\documentclass[%
 reprint,
superscriptaddress,
nobibnotes,
 amsmath,amssymb,
 aps,
prl 
]{revtex4-2}

\usepackage{aas_macros}
\usepackage{graphicx}%
\graphicspath{{./}{figures/}}
\usepackage{dcolumn}%
\usepackage{bm}%
\usepackage[usenames,dvipsnames]{xcolor}
\usepackage{hyperref}%
\hypersetup{
	colorlinks=true,
	citecolor=NavyBlue,
	linkcolor=NavyBlue,
	urlcolor=NavyBlue,
}

\usepackage{epsfig}
\usepackage{url}
\usepackage[normalem]{ulem}
\usepackage{latexsym}
\usepackage{epsfig}
\usepackage{amsmath}
\usepackage{amssymb}
\usepackage{wasysym}
\usepackage{graphicx}
\usepackage{verbatim}
\usepackage{enumerate,mdwlist}%
\usepackage[titletoc]{appendix}
\usepackage{amsfonts}
\usepackage{mathrsfs}
\usepackage{pgfplots}
\usepackage[export]{adjustbox}
\usepackage{bbold}

\newcommand{\lettersection}[1]{\textbf{\emph{#1}}.---}
\newcommand{\this}{paper}

\newcommand{\rs}{r_{*}}
\newcommand{\rp}{r_{+}}

\newcommand{\diff}{d}
\newcommand{\hor}{\mathrm{H}}

\newcommand{\GSNCode}{\texttt{GeneralizedSasakiNakamura.jl}}

\newcommand{\SM}{Supplemental Material \cite{SM}}

\usepackage[acronym]{glossaries}
\newacronym{GSN}{GSN}{generalized Sasaki-Nakamura}
\newacronym{SN}{SN}{Sasaki-Nakamura}
\newacronym{MST}{MST}{Mano-Suzuki-Takasugi}
\newacronym{NP}{NP}{Newman-Penrose}
\newacronym{BH}{BH}{black hole}
\newacronym{GW}{GW}{gravitational wave}
\newacronym{ODE}{ODE}{ordinary differential equation}
\newacronym{PDE}{PDE}{partial differential equation}
\newacronym{GDT}{GDT}{Generalized Darboux Transformation}
\newacronym{EMRI}{EMRI}{extreme mass-ratio inspiral}
\newacronym{LIGO}{LIGO}{Laser Interferometer Gravitational-Wave Observatory}
\newacronym{LISA}{LISA}{Laser Interferometer Space Antenna}
\newacronym{DECIGO}{DECIGO}{Deci-hertz Interferometer Gravitational wave Observatory}
\newacronym{AD}{AD}{automatic differentiation}
\newacronym{IBP}{IBP}{integration by parts}
\newacronym{SWSH}{SWSH}{spin-weighted spheroidal harmonic}
\newacronym{BL}{BL}{Boyer-Lindquist}
\newacronym{FT}{FT}{Fourier transform}
\newacronym{ZFL}{ZFL}{zero frequency limit}
\newacronym{QNM}{QNM}{quasinormal mode}
\newacronym{RW}{RW}{Regge-Wheeler}
\newacronym{ECO}{ECO}{exotic compact object}
\newacronym{TS}{TS}{Teukolsky-Starobinsky}

\makeatletter
\newcommand*{\glsplainhyperlink}[2]{%
  \colorlet{currenttext}{.}%
  \colorlet{currentlink}{\@linkcolor}%
  \hypersetup{linkcolor=currenttext}%
  \hyperlink{#1}{#2}%
  \hypersetup{linkcolor=currentlink}%
}
\let\@glslink\glsplainhyperlink
\makeatother

\begin{document}

\title{Near-horizon gravitational perturbations of rotating black holes}
\author{Rico K.~L. Lo}
\email{kalok.lo@nbi.ku.dk}
\affiliation{%
Center of Gravity, Niels Bohr Institute, Blegdamsvej 17, 2100 Copenhagen, Denmark}%
\author{Yucheng Yin}
\email{yucheng.yin@nbi.ku.dk}
\affiliation{%
Center of Gravity, Niels Bohr Institute, Blegdamsvej 17, 2100 Copenhagen, Denmark}%
\affiliation{Department of Astronomy, School of Physics, Peking University, 100871 Beijing, China}%
\affiliation{Kavli Institute for Astronomy and Astrophysics at Peking University, 100871 Beijing, China}%

\date{\today}%

\begin{abstract}
Perturbative calculations of gravitational radiation near the horizons of rotating black holes in the frequency domain have been plagued by divergence issues.
We resolve this longstanding obstacle by constructing a nonsingular source term for near-horizon gravitational perturbations, or equivalently perturbed Weyl scalars $\psi_0$ with a spin weight of $s = +2$, within the generalized Sasaki-Nakamura formalism for the first time.
As illustrative applications, we compute the dynamical deformation of the event horizon induced by an ultrarelativistic particle plunge, demonstrating the excitation of quasinormal modes at the horizon, and we evaluate the energy flux toward the horizon from an extreme mass-ratio inspiral.
This work provides a powerful tool for studying physics near black hole horizons.
\end{abstract}

\maketitle

\glsresetall
\lettersection{Introduction}%
Astrophysical \glspl{BH} are rarely isolated and constantly interact with their surroundings gravitationally.
For instance, they can be found at the center of galaxies where those supermassive \glspl{BH} are surrounded by much lighter objects such as stars and stellar-mass \glspl{BH} \cite{Magorrian:1997hw, Ghez_1998, Ghez:2008ms}.
When these nearby objects orbit close enough to the central \gls{BH}, they spiral more into the \gls{BH} as they lose energy through the emissions of \glspl{GW} that are referred to as \glspl{EMRI} \cite{2018LRR....21....4A}.
These \gls{GW} signals can be detected by future space-based detectors such as the \gls{LISA} \cite{Amaro-Seoane:2007osp, Gair:2017ynp, LISAConsortiumWaveformWorkingGroup:2023arg}.
\Glspl{BH} are also found in pairs, where the mergers of these binary \glspl{BH} produced some of the most energetic astronomical signals ever observed, by the ground-based LIGO-Virgo-KAGRA detector network \cite{LIGOScientific:2016aoc, LIGOScientific:2025rid}.
After a merger, the remnant object relaxes to a stationary state by emitting \glspl{GW} with characteristic frequencies, a process known as the ringdown \cite{Berti:2025hly}.

These phenomena can be studied using \gls{BH} perturbation theory \cite{Sasaki:2003xr, Pound:2021qin}, where a perturbed spacetime (e.g., a supermassive \gls{BH} encircled by a stellar-mass \gls{BH}) is expressed as a small perturbation on top of an analytically known background metric.
By linearizing the Einstein field equations, one can find the set of linear \glspl{PDE} and source terms that describe how a perturber affects the spacetime.
Solutions to these equations give us, for example, the waveforms of \gls{GW} signals that we observe at the detectors.
Comparing these theoretical predictions with observation data then allows us to understand more about the source of those \glspl{GW} \cite{LIGOScientific:2025slb}. 

\lettersection{Perturbations of Kerr black holes using the Teukolsky formalism}%
The Kerr metric \cite{Kerr:1963ud} is an exact solution that describes an isolated spinning \gls{BH} of mass $M$, angular momentum $aM$ and no electric charge. Using the Boyer-Lindquist coordinates $(t,r,\theta,\varphi)$ \cite{Boyer:1966qh}, its line element is given by
\begin{multline}
\label{eq:Kerr_metric}
	ds^2 = 	-\left(1 - \dfrac{2r}{\Sigma} \right) dt^2 - \dfrac{4ar\sin^2 \theta}{\Sigma} dt d\varphi + \dfrac{\Sigma}{\Delta} dr^2 \\
	+ \Sigma d\theta^2 + \sin^2 \theta \left( r^2 + a^2 + \dfrac{2a^2 r\sin^2 \theta}{\Sigma} \right) d\varphi^2,
\end{multline}
where $\Sigma \equiv r^2 + a^2 \cos^2 \theta$ and $\Delta \equiv r^2 - 2r + a^2 = (r - r_{+})(r - r_{-})$ with $r_{\pm} = 1 \pm \sqrt{1 - a^2}$ being the outer and the inner event horizon, respectively \footnote{Throughout the paper, we use geometric units where $c = G = M = 1$.}.
With this as the background metric, we can now in principle perform the linearization and compute the perturbation to the metric tensor.

Instead of computing directly the tensorial metric perturbations, Teukolsky showed that it is much easier to calculate instead the perturbations to scalar projections of the Weyl curvature tensor, referred to as Weyl scalars \cite{Teukolsky:1973ha}.
Furthermore, this enables for the separation of variables on the governing \glspl{PDE}, despite the lack of spherical symmetry as seen in Eq.~\eqref{eq:Kerr_metric}.
In the Teukolsky formalism, we work with a radial function ${}_{s}R_{\ell m \omega}(r)$ and an angular function ${}_{s}S_{\ell m \omega}(\theta, \varphi)$, satisfying the radial and the angular Teukolsky equation, respectively, which are \glspl{ODE} instead.
The subscript $\omega$ denotes the Fourier variable (or physically the angular frequency), while $(\ell, m)$ label the discrete eigenfunctions in the angular sector that satisfy the regular boundary conditions at $\theta = 0, \pi$ while respecting the azimuthal symmetry, similar to spherical harmonics \footnote{In fact, solutions to the angular Teukolsky equation ${}_{s}S_{\ell m \omega}(\theta, \phi)$ are known as spin-weighted spheroidal harmonics.}.

The subscript $s$ denotes the spin weight of the perturbation, which also indicates the Weyl scalar that is being perturbed.
Gravitational perturbations have $s = \pm 2$.
However, one is usually interested in the $s = -2$ case only, which corresponds to the Weyl scalar $\psi_4$.
This is because $\psi_4$ encodes, for example, the waveform and the energy flux of the outgoing gravitational radiation at future null infinity that we observe at \gls{GW} detectors \cite{Teukolsky:1973ha}.

Near a \gls{BH} horizon, information about the ingoing gravitational radiation, such as the energy flux toward the horizon, is encoded in the Weyl scalar $\psi_0$ instead, which corresponds to the $s = +2$ case \cite{Teukolsky:1973ha}.
While we cannot directly measure the gravitational radiation near a \gls{BH}, it is still observationally relevant.
This is because we need to compute \emph{both} fluxes toward infinity and the \gls{BH} horizon when incorporating the radiation reaction due to the emission of \gls{GW} in waveform modeling.
There is also a growing interest in the community on studying the radiation falling toward the horizon.
For example, there was a recent proposal of ``\gls{BH} tomography'' where one attempts to map the geometry of the horizon using observations made at infinity \cite{Ashtekar:2021kqj, RibesMetidieri:2024tpk, RibesMetidieri:2025lxr}.
The absence of a horizon, as proposed in some models for \acrlongpl{ECO}, can also leave imprints referred to as \gls{GW} echoes, on the gravitational waveforms we detect \cite{Cardoso:2016rao, Xin:2021zir, Srivastava:2021uku}.
All of these require us to solve for gravitational perturbations near horizons (or would-be horizons).

For both $s = \pm 2$ cases, the inhomogeneous radial Teukolsky equation has the same form. In the frequency domain, the equation is given by
\begin{equation}
\label{eq:inhomogeneous_radialteukolsky}
	\left[\Delta^{-s} \dfrac{\diff}{\diff r}\left( \Delta^{s+1} \dfrac{\diff }{\diff r} \right) - V_{\rm T}(r)\right] {}_{s}R_{\ell m \omega} = - {}_{s}\mathcal{T}_{\ell m \omega}(r),
\end{equation}
where $V_{\rm T}$ is referred to as the potential of the Teukolsky equation \footnote{An expression using notations and conventions consistent with this paper can be found in Ref.~\cite{Lo:2023fvv}.}, and ${}_{s}\mathcal{T}_{\ell m \omega}$ is the source term of the equation.
A solution satisfying the boundary condition of being purely ingoing at the horizon and purely outgoing at infinity can be constructed using the Green's function method \cite{ARFKEN2013447}, which is given by
\begin{multline}
\label{eq:R_inhomo}
	{}_{s}R^{\rm inhomo}_{\ell m \omega}(r) =  \frac{{}_{s}R^{\rm up}_{\ell m \omega}(r)}{W_R} \int_{r_+}^{r} \dfrac{{}_{s}R^{\rm in}_{\ell m \omega}(\tilde{r}) {}_{s}\mathcal{T}_{\ell m \omega}(\tilde{r})}{\Delta^{-s}(\tilde{r})} d\tilde{r} \\
	 + \frac{{}_{s}R^{\rm in}_{\ell m \omega}(r)}{W_R} \int_{r}^{\infty} \dfrac{{}_{s}R^{\rm up}_{\ell m \omega}(\tilde{r}) {}_{s}\mathcal{T}_{\ell m \omega}(\tilde{r})}{\Delta^{-s}(\tilde{r})} d\tilde{r},
\end{multline}
where $W_{R}$ is the Wronskian of the two homogeneous solutions ${}_{s}R^{\rm in, up}_{\ell m \omega}$ that satisfy the boundary condition at the horizon and infinity, respectively \cite{Lo:2023fvv}.

Unfortunately, \emph{this} Green's function solution requires some regularization when the source term ${}_{s}\mathcal{T}_{\ell m \omega}$ is nonvanishing at spatial infinity for $s = -2$ \cite{Poisson:1996ya, Campanelli:1997sg} or near the horizon for $s = +2$ \cite{Srivastava:2021uku}.
For example, ${}_{-2}\mathcal{T}_{\ell m \omega}(r \to \infty)$ diverges as $r^{7/2}$ and ${}_{+2}\mathcal{T}_{\ell m \omega}(r \to \rp)$ diverges as $1/\left(r - \rp\right)^2$ for a particle falling initially at rest from infinity to the horizon of a \gls{BH} radially.
These regularization procedures work in principle but they are cumbersome to execute in practice.

This divergence issue of the source term for $s = -2$ led to the development of the \gls{SN} formalism \cite{SASAKI198268, 10.1143/PTP.67.1788}, which is designed to give a nonsingular source term when solving the inhomogeneous equation using the Green's function method.
Indeed, the \gls{SN} formalism has found great success in \gls{BH} perturbation theory calculations (for instance, Refs.~\cite{Oshita:2021iyn, Oshita:2022yry, Watarai:2024vni, Watarai:2024huy, Oshita:2025qmn}).

The same, however, cannot be said for $s = +2$.
The \gls{GSN} formalism---first developed by Hughes \cite{Hughes:2000pf}---had limited applications thus far, only for solving the homogeneous Teukolsky equations (when ${}_{s}\mathcal{T}_{\ell m \omega} = 0$) with integer spin weights $s = 0, \pm 1, \pm 2$ \cite{Lo:2023fvv, Lo:2025njp}.
This is primarily due to the lack of a source term for the inhomogeneous \gls{GSN} equation.
In this \this{}, we fill in this void and for the first time present the source term for $s = +2$ or equivalently $\psi_0$ perturbations, and show that it is finite near a \gls{BH} horizon.

\lettersection{Gravitational perturbations of Kerr black holes using the GSN formalism}%
In place of the radial Teukolsky equation in Eq.~\eqref{eq:inhomogeneous_radialteukolsky}, we solve the inhomogeneous \gls{GSN} equation for the solution ${}_{s}X_{\ell m \omega}$. The equation reads
\begin{equation}
\label{eq:inhomo_GSNeqn}
    \left[ \dfrac{\diff^2}{\diff \rs^2} - {}_{s}\mathcal{F}_{\ell m \omega}(r) \dfrac{\diff}{\diff \rs} - {}_{s}\mathcal{U}_{\ell m \omega}(r) \right] {}_{s}X_{\ell m \omega} = {}_{s}\mathcal{S}_{\ell m \omega}(r),
\end{equation}
where ${}_{s}\mathcal{S}_{\ell m \omega}$ is the \gls{GSN} source term and $\rs$ is the tortoise coordinate which maps the \gls{BH} horizon at $r = \rp$ to $\rs = -\infty$ while keeping spatial infinity at $\rs = \infty$.
The expressions for ${}_{s}\mathcal{F}_{\ell m \omega}$ and ${}_{s}\mathcal{U}_{\ell m \omega}$ can be found in Ref.~\cite{Lo:2023fvv}.
After ${}_{s}X_{\ell m \omega}$ has been computed, one can obtain also ${}_{s}R_{\ell m \omega}$.
Specifically, for gravitational perturbations with $s = +2$, we have
\begin{equation}
	{}_{+2}R_{\ell m \omega}(r) = {}_{+2}\Lambda^{-1}_{\ell m \omega}\left( {}_{+2}X_{\ell m \omega} \right) + \dfrac{1}{\eta} \dfrac{\left(r^2 + a^2\right)^{3/2}}{\Delta^{2 }} {}_{+2}\mathcal{S}_{\ell m \omega},
\end{equation}
with ${}_{+ 2}\Lambda^{-1}_{\ell m \omega}$ being the inverse operator that transforms a solution to the homogeneous \gls{GSN} equation (when ${}_{s}\mathcal{S}_{\ell m \omega} = 0$) to the corresponding homogeneous Teukolsky solution. The expressions for the $\Lambda^{-1}$ operator \footnote{We omit the $s$ subscript (on the left) and the $\ell m \omega$ subscripts (on the right) unless there is an ambiguity or we want to emphasize the dependence on them.} and $\eta(r)$ can both be found in Ref.~\cite{Lo:2023fvv}.
The key result here---the \gls{GSN} source term ${_{+2}}\mathcal{S}_{\ell m \omega}$ for $s = +2$---is given by
\begin{equation}
\label{eq:calS}
	{_{+2}}\mathcal{S}_{\ell m \omega} = \dfrac{\eta \Delta}{ (r^2 + a^2)^{3/2} r^2}  {_{+2}}\mathcal{W}_{\ell m \omega} \exp \left( \int^{r} i\dfrac{K}{\Delta} d\tilde{r} \right),
\end{equation}
where ${_{+2}}\mathcal{W}_{\ell m \omega}$ satisfies
\begin{equation}
\label{eq:ODE_for_calW}
    \dfrac{\diff^2 {_{+2}}\mathcal{W}_{\ell m \omega}}{\diff r^2} = - r^2 {}_{+2}\mathcal{T}_{\ell m \omega}  \exp \left( -\int^{r} i\dfrac{K}{\Delta} d\tilde{r} \right).
\end{equation}
The derivation of the newly constructed source term, which is valid for \emph{arbitrary} motions (not limited only to geodesics) of a point-particle perturber, is provided in the \SM{} \footnote{For $s = -2$, the expressions and their derivation can be found in Appendix A of Ref.~\cite{Yin:2025kls}.}.

Just like the radial Teukolsky equation, an inhomogeneous \gls{GSN} solution satisfying the same boundary conditions as Eq.~\eqref{eq:R_inhomo} can be constructed using the Green's function method.
Specifically, near the \gls{BH} horizon, the solution ${}_{+2}X^{\rm inhomo}_{\ell m \omega}(\rs \to -\infty)$ can be written as
\begin{equation}
\label{eq:full_inhomo_X_using_green_func}
    \begin{aligned}
        & {}_{+2}X^{\rm inhomo}_{\ell m\omega}(\rs \to -\infty) \\
        & = \frac{{}_{+2}X_{\ell m\omega}^{\mathrm{in}}(\rs \to -\infty)}{W_X} \int_{-\infty}^{\infty} {}_{+2}X_{\ell m\omega}^{\mathrm{up}}(\tilde{r}_*)\frac{{}_{+2}\mathcal{S}_{\ell m\omega}(\tilde{r}_*)}{\eta}\diff\tilde{r}_* \\
        & = \underbrace{\left[ \dfrac{B^{\rm trans}_{\rm SN}}{W_{X}} \int_{-\infty}^{\infty} {}_{+2}X_{\ell m\omega}^{\mathrm{up}}(\tilde{r}_*)\frac{{}_{+2}\mathcal{S}_{\ell m\omega}(\tilde{r}_*)}{\eta}\diff\tilde{r}_* \right]}_{{}_{+2}X^{\hor}_{\ell m \omega}} e^{-i\kappa \rs},
    \end{aligned}
\end{equation}
where $\kappa \equiv \omega-ma/(2\rp)$ is the effective wave frequency at the horizon and $W_{X}$ is the Wronskian of the two homogeneous solutions ${}_{s}X^{\rm in, up}_{\ell m \omega}$ \cite{Lo:2023fvv}.
Here, we use the fact that asymptotically $X^{\mathrm{in}}(\rs \to -\infty) = B^{\rm trans}_{\rm SN} e^{-i\kappa \rs}$.

Crucially, the integral in Eq.~\eqref{eq:full_inhomo_X_using_green_func} for ${}_{+2}X^{\hor}_{\ell m \omega}$ is \emph{absolutely convergent} as $\rs \to -\infty$ when approaching the \gls{BH} horizon, unlike the corresponding integral in Eq.~\eqref{eq:R_inhomo} that is divergent for the Teukolsky formalism \footnote{A detailed analysis can be found in the \SM{}.}.
Our construction of the source term for $s = +2$ in the \gls{GSN} formalism, therefore, enables calculations that were previously impossible in the frequency domain using the Teukolsky formalism without regularization.

\lettersection{Radial infalls}%
To demonstrate the capability of the newly extended \gls{GSN} formalism, here we consider a test particle of rest mass $\mu \ll M$ falling radially from infinity into a nonrotating \gls{BH} of mass $M$ with an initial speed close to the speed of light, such that the energy per unit rest mass (equivalently the Lorentz boost factor) $\mathcal{E} \gg 1$.
Without loss of generality, we set the trajectory as $\theta(t) = \varphi(t) = 0 \; \forall t \in \left( -\infty, +\infty \right)$ and consider modes with $m =  0$ only.
With this setup, an analytical solution for $\mathcal{W}$ and hence the \gls{GSN} source term $\mathcal{S}$ can be obtained.

The shear $\sigma$ encodes the deformation of the now-perturbed horizon \cite{Poisson:2004cw, OSullivan:2014ywd, OSullivan:2015lni}.
Using the Hawking-Hartle tetrad, it is related to ${}_{+2}X^{\hor}_{\ell m \omega}$ by \cite{Hawking:1972hy, Teukolsky:1974yv}
\begin{multline}
    \sigma(r \to \rp) = \frac{1}{4(\rp^2+a^2)^2} \int_{-\infty}^{\infty} d\omega \frac{i}{\kappa-2i\epsilon_0} \dfrac{B^{\rm trans}_{\rm T}}{B^{\rm trans}_{\rm SN}} \\
      \left[ {}_{+2}X^{\hor}_{\ell m \omega} e^{i\left(\frac{ma}{2\rp}\right)\rs} {}_{+2}S_{\ell m \omega}(\Theta, \Phi) e^{-i\omega v}\right],
\end{multline}
where the conversion factor $B^{\rm trans}_{\rm T}/B^{\rm trans}_{\rm SN}$ can be found in Ref.~\cite{Lo:2023fvv}, $\epsilon_0=\sqrt{1-a^2}/(4\rp)$, $v \equiv t  + \rs$ is the advanced time and $(\Theta, \Phi)$ are the viewing angles \footnote{Moreover, it is assumed that the perturbed horizon is still located at $r = \rp$. This is justified since one can always choose a ``horizon-locking'' gauge such that this assumption holds true \cite{Poisson:2004cw, Stein:2024yxc}.}.
The same calculation can also be done in the Teukolsky formalism with the aid of the regularization scheme in Ref.~\cite{Srivastava:2021uku}.
The details of both calculations (such as the expression of the analytical \gls{GSN} source term) can be found in the \SM{}.
Note that the \gls{TS} identities \cite{Starobinskil:1974nkd, Teukolsky:1974yv} cannot be used to convert a $s = -2$ calculation into a $s = +2$ one that we need here, since the Teukolsky source term is nonzero at the horizon (see also, for example, Fig.~4 in Ref.~\cite{Srivastava:2021uku}).

\begin{figure}[htpb!]
    \centering
    \includegraphics[width=1.0\linewidth]{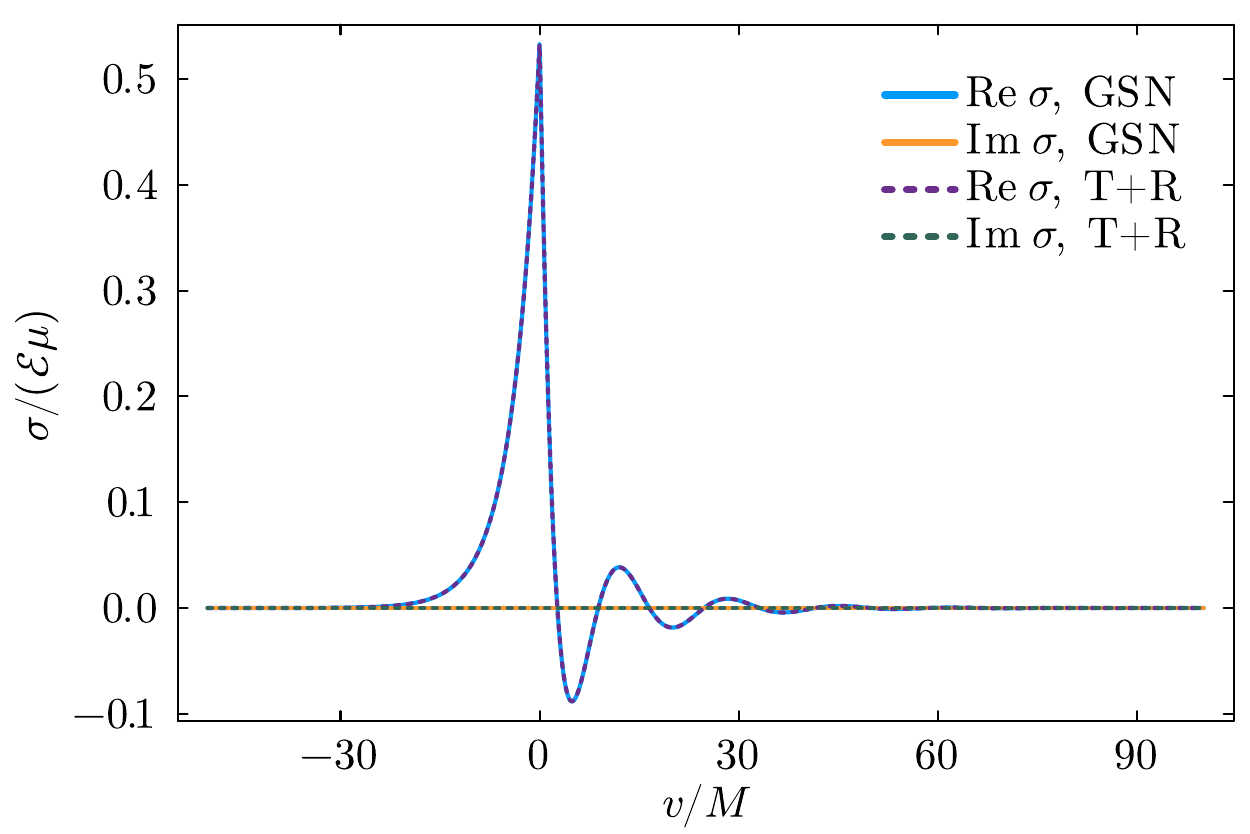}
    \caption{The induced waveform of the perturbed shear for the $\ell = 2$ mode at the horizon $\sigma(r \to \rp)$ by an ultrarelativistic particle falling radially into a nonrotating \gls{BH}, viewing at $\Theta=\pi/2$, $\Phi=0$. The \gls{GSN} and {Teukolsky+regularization} (T+R) approaches agree very well.}
    \label{fig:Waveform_hor}
\end{figure}

Figure~\ref{fig:Waveform_hor} shows the shear at the \gls{BH} horizon induced by the radial infall of the ultrarelativistic particle for the $\ell = 2$ mode as a function of $v$.
The two calculations agree well up to numerical errors, with a maximum difference $\sim 10^{-8}$. 
Meanwhile, the \gls{GSN} approach is $\sim 18$ times faster than the {Teukolsky+regularization} approach \footnote{In this numerical experiment, the homogeneous Teukolsky solutions were transformed from the homogeneous \gls{GSN} solutions, where the transformation takes a negligible amount of time, in order to ensure a fair comparison. Both calculations were done on the same machine.}.
This speedup is purely due to the \emph{superior convergence} of the Green's function integral in the \gls{GSN} formalism.

The particle crosses the \gls{BH} horizon at $v = 0$.
Visually from Fig.~\ref{fig:Waveform_hor}, we see that the shear perturbation ``rings down'' for $v > 0$.
This suggests that \glspl{QNM} are excited.
Physically, it is hardly surprising that the \gls{BH} oscillates at those characteristic frequencies to get rid of the perturbation.
In fact, the excitation of \glspl{QNM} at the horizon has been reported in literature studying numerical relativity simulations \cite{Mourier:2020mwa, Khera:2023oyf}.

\begin{figure}[htpb!]
    \centering
    \includegraphics[width=1.0\linewidth]{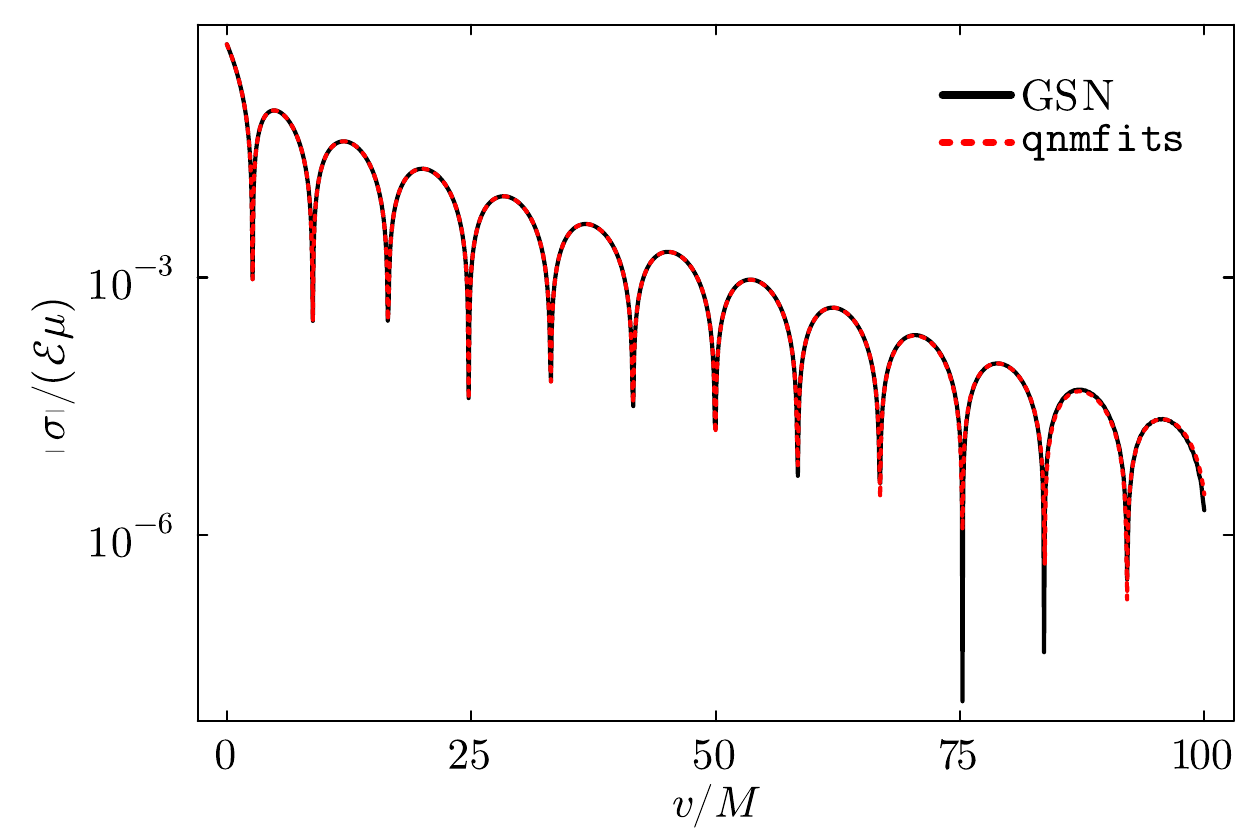}
    \caption{A fit using gravitational \glspl{QNM} (dashed) on the shear waveform in Fig.~\ref{fig:Waveform_hor} (solid). The fit uses $\ell = 2$ and $n = 0 \dots 3$ overtones, achieving a mismatch of $1.22\times 10^{-8}$.}
    \label{fig:QNM_fit}
\end{figure}

We confirm that \glspl{QNM} are indeed excited by performing a fit with the gravitational \glspl{QNM} for a nonrotating \gls{BH} (with $\ell = 2$) on the shear waveform in Fig.~\ref{fig:Waveform_hor}, from $v = 0M$ to $v = 100M$ using \texttt{qnmfits} \cite{MaganaZertuche:2025bua}.
A fit of the waveform that includes up to the third overtone is shown in Fig.~\ref{fig:QNM_fit}, where it achieves a mismatch $\mathcal{M}$, defined in Ref.~\cite{MaganaZertuche:2025bua}, of $1.22\times 10^{-8}$.
Note that the mass and the spin parameter are fixed to $M = 1$ and $a/M = 0$, respectively.
The free parameters during the fitting are therefore only the amplitudes for each of the \glspl{QNM}.
As a simple check, we repeat the fitting including up to the $n_{\rm max}$th overtone, for $n_{\rm max} = 0$ to $n_{\rm max} = 7$.
The values for the mismatch $\mathcal{M}$ as a function of $n_{\rm max}$ are tabulated in Table~\ref{tab:Mismatch}.
Despite the $n_{\rm max} = 7$ fit having the lowest mismatch among those tested, we find that we cannot reliably extract the \gls{QNM} amplitudes beyond the third overtone \footnote{These kind of behaviors are not surprising and have been reported in, for example, Refs.~\cite{Forteza:2021wfq, Baibhav:2023clw}.}.
In addition, we find that the fundamental mode alone (i.e., $n_{\rm max} = 0$) is not enough to fit the shear waveform, and more overtones are necessary to extract its amplitude robustly (refer to the \SM{} for more details).

\begin{table}[h]
    \caption{Mismatch $\mathcal{M}$ as a function of $n_{\mathrm{max}}$, the number of overtones included in the fit.}
    \label{tab:Mismatch}
    \centering
    \begin{ruledtabular}
    \begin{tabular}{cccc}
    $n_{\rm max}$ & $\mathcal{M}$ & $n_{\rm max}$ & $\mathcal{M}$ \\
    \hline
    $0$& $2.58\times 10^{-1}$ & $4$ & $7.07\times 10^{-9}$ \\ 
    $1$ & $6.15\times 10^{-3}$ & $5$ & $6.27\times 10^{-9}$ \\
    $2$ & $1.36\times 10^{-5}$ & $6$ & $4.90\times 10^{-9}$ \\
    $3$ & $1.22\times 10^{-8}$ & $7$ & $3.32\times 10^{-9}$ \\
    \end{tabular}
    \end{ruledtabular}
\end{table}

\lettersection{Horizon fluxes for extreme mass-ratio inspirals}%
Another application of the \gls{GSN} formalism is to compute near-horizon gravitational perturbations due to \glspl{EMRI}.
Most literature, such as Refs.~\cite{OSullivan:2014ywd, OSullivan:2015lni}, uses the \gls{TS} identities \cite{Starobinskil:1974nkd, Teukolsky:1974yv} to convert $\psi_4$ with a $s = -2$ source term into $\psi_0$.
This is valid because the Teukolsky source term vanishes at the \gls{BH} horizon for these bound inspiral orbits.
Here with the \gls{GSN} formalism, we compute \emph{directly} $\psi_0$ with a $s = +2$ source term and then its derived quantities such as the energy flux down the horizon.

Instead of numerically integrating Eq.~\eqref{eq:ODE_for_calW} for the \gls{GSN} source term, we use the \gls{IBP}-based scheme in Ref.~\cite{Yin:2025kls} to bypass the extra radial integration.
Since the \gls{GSN} source term $\mathcal{S}$ for $s = +2$ has the same form as the one for $s = -2$ except for a swap in the sign inside the exponential function, it means that we can reuse the same construction of two new auxiliary functions ${}_{+2}Y^{\rm in, up}$.
More explicitly, they are given by
\begin{equation}\label{Eq.ODEforY}
    {}_{+2}Y_{\ell m\omega}^{\rm in/up\ \prime\prime}(r)\equiv \frac{{}_{+2}X_{\ell m\omega}^{\rm in/up}(r)}{r^2\sqrt{r^2+a^2}}\exp\left(i\int^r\frac{K}{\Delta}\diff r\right).
\end{equation}
For generic bound and stable orbits, only a discrete set of frequencies $\omega$ contribute to the spectrum, i.e., $\omega \to \left\{n, k\right\}$ where $n, k$ are integers referred to as the radial and the polar index, respectively.
Specifically, we have, per $(\ell, m, n, k)$ mode, that
\begin{equation}
\label{eq:XH_lmnk}
    {}_{+2}X^{\hor}_{\ell mnk} = \dfrac{B^{\rm trans}_{\rm SN}}{W_{X}} \int_{\rp}^{\infty} {}_{+2}Y^{\rm up}_{\ell m \omega}(r) \dfrac{\diff^2 {}_{+2}\mathcal{W}_{\ell m \omega}}{\diff r^2} dr,
\end{equation}
since all of the boundary terms vanish if we choose the boundary conditions for ${}_{+2}Y_{\ell m \omega}^{\rm up}$ as $Y(r\to\infty)=Y'(r\to\infty)=0$ \cite{Yin:2025kls}.

The averaged energy flux per unit rest mass toward the horizon $\left\langle \diff{\mathcal{E}}/\diff t \right\rangle^{\hor}$, for example, is related to ${}_{+2}X^{\hor}_{\ell mnk}$ by \cite{Chandrasekhar:1985kt}
\begin{equation}\label{Eq.EnergyFlux}
    \left\langle \diff{\mathcal{E}}/\diff t \right\rangle^{\hor}_{\ell mnk}=\frac{\omega\left|{\left(B^{\rm trans}_{\rm T}/B^{\rm trans}_{\rm SN}\right)} {}_{+2}X^{\hor}_{\ell mnk}\right|^2}{64\pi(2\rp)^3\kappa(\kappa^2+4\epsilon_0^2)}.
\end{equation}
Similar expressions for the angular momentum flux and the Carter constant flux toward the \gls{BH} horizon can be found in the \SM{}.

\begin{figure}[htpb!]
    \centering
    \includegraphics[width=1.0\linewidth]{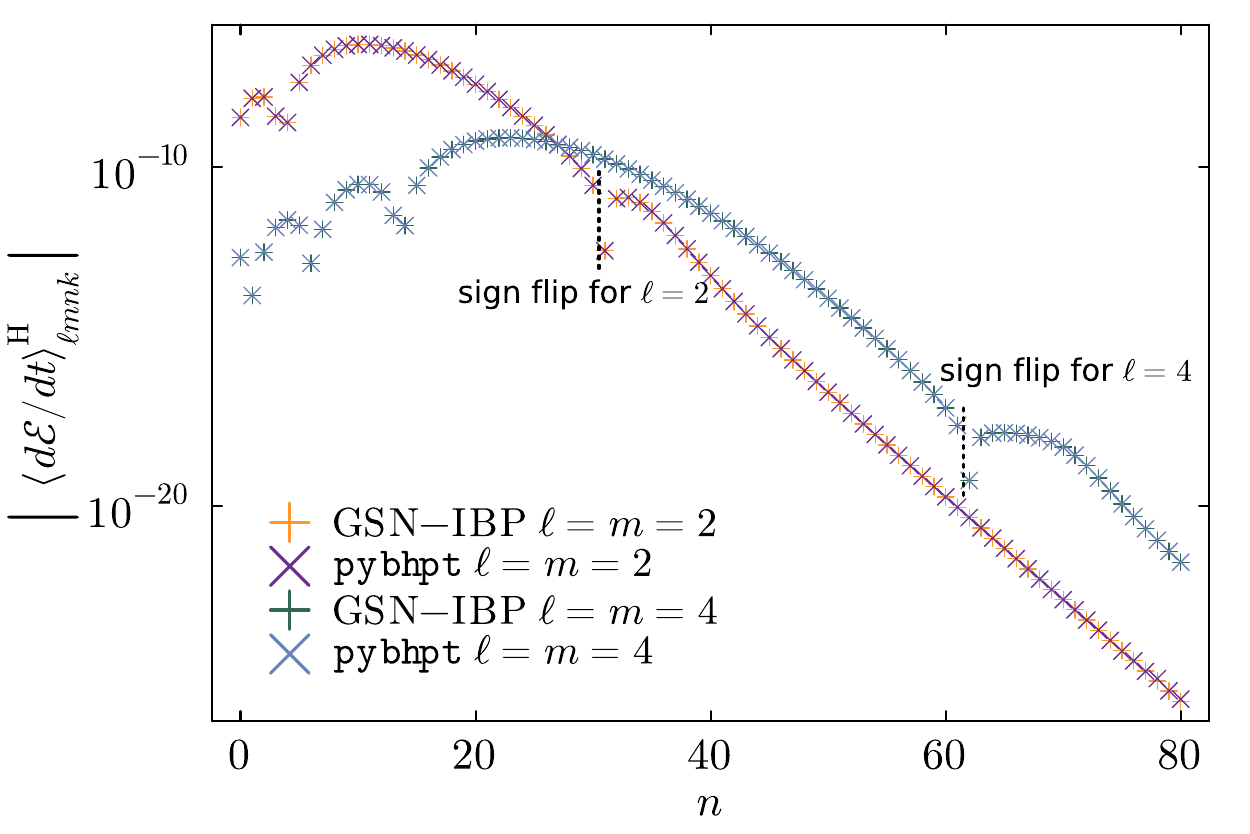}
    \caption{The absolute value of energy flux toward the \gls{BH} horizon for $a=0.9M$, $p=6M$, $e=0.7$, $x=\cos\pi/4$. The mode indexes are $\ell=m=2$ and $\ell=m=4$ with polar index $k=0$ and radial index $n=0$ to $n=80$. The two dashed lines mark the transition of the energy flux from being negative to positive due to the sign change of $\kappa$. For $\ell=2$, this transition occurs between $n=30$ and $n=31$; for $\ell=4$, it takes place between $n=61$ and $n=62$. The \gls{GSN}-\gls{IBP} and Teukolsky (using \texttt{pybhpt}) approaches agree very well.}
    \label{fig:bound_flux_hor}
\end{figure}

Figure \ref{fig:bound_flux_hor} shows the absolute value $\left|\left\langle \diff{\mathcal{E}}/\diff t \right\rangle^{\hor}_{\ell m n k} \right|$ for a set of fiducial parameters ($a=0.9M$, $p=6M$, $e=0.7$, $x=\cos\pi/4$, for the spin parameter, semilatus rectum, eccentricity and inclination parameter, respectively), and $n = 0 \dots 80$, $k = 0$ with $\ell = m = 2$ and $\ell = m = 4$, calculated using our \gls{GSN}-\gls{IBP} approach and the Teukolsky formalism as implemented in \texttt{pybhpt} \cite{PhysRevD.106.064042, PhysRevD.109.044020} (using the \gls{TS} identities), respectively.
The two set of results agree very well as expected.
Table \ref{tab:horizon_flux_values} tabulates the values of $\langle \diff{\mathcal{E}}/\diff t \rangle^{\rm H}_\ell$ (i.e., summed over $m,n,k$) using the same set of fiducial parameters above, for $\ell = 2 \dots 6$, using both the \gls{GSN}-\gls{IBP} approach and \texttt{pybhpt}.
We adopted the same truncation rules for the summation as in Ref.~\cite{Yin:2025kls}.
The two set of results agree to the sixteenth digit.
All these are now implemented as part of the \GSNCode{} code \footnote{\url{https://github.com/ricokaloklo/GeneralizedSasakiNakamura.jl} from v0.8.0 onward.}.
\begin{table}[h]
    \caption{The energy fluxes of different $\ell$ modes for $a=0.9M$, $p=6M$, $e=0.7$, $x=\cos\pi/4$ toward the \gls{BH} horizon. The last column is the total number of modes in the summation. The two agree up to the sixteenth digit.}
    \label{tab:horizon_flux_values}
    \centering
    \begin{ruledtabular}
    \begin{tabular}{cccc}
    $\langle \diff{\mathcal{E}}/\diff t \rangle^{\rm H}_\ell$ & \gls{GSN}-\gls{IBP}\ $(\times 10^{-5})$ & \texttt{pybhpt}\ $(\times 10^{-5})$ & Modes\\ 
    \hline
    $\ell=2$ & $-1.05673338761(886)$ & $-1.05673338761(892)$ & $1194$ \\
    $\ell=3$ & $-0.07124204965(416)$ & $-0.07124204965(417)$ & $1748$ \\
    $\ell=4$ & $-0.00612365199(756)$ & $-0.00612365199(756)$ & $2038$ \\
    $\ell=5$ & $-0.00056415647(036)$ & $-0.00056415647(036)$ & $2740$ \\
    $\ell=6$ & $-0.00004527861(683)$ & $-0.00004527861(683)$ & $2748$ \\
    \end{tabular}
    \end{ruledtabular}
\end{table}

\lettersection{Conclusion}%
We present here, for the first time, the source term for gravitational perturbations under the \gls{GSN} formalism to compute gravitational radiation near the horizon of a Kerr \gls{BH}.
Importantly, with this formalism, the convolution integral involved when solving the inhomogeneous wave equation with the Green's function method in the frequency domain is now convergent, and hence there is no need for any regularization, which is crucial for studying horizon physics with perturbative calculations.

We demonstrate the capability of our new approach with two cases---the radial infall of an ultrarelativistic particle into a nonrotating \gls{BH} and the energy flux of an \gls{EMRI} down the horizon of a rotating \gls{BH}.
Specifically for the radial infall, we show that \glspl{QNM} are excited at the horizon, and they provide a good fit of the shear waveform when sufficient number of overtones are included.

Our new tool has a broad range of applications, such as gravitational physics and astrophysics.
For example, one can now perform perturbative computations of particles plunging into a \gls{BH} with various trajectories to study the excitation of \glspl{QNM} at the \gls{BH} horizon in-depth.
We envision that the \gls{GSN} formalism can be further developed for applications such as metric reconstruction and second-order perturbation calculations, which are relevant for constructing waveform templates for \gls{GW} detectors like \gls{LISA}.

This work, together with Refs.~\cite{Lo:2023fvv, Lo:2025njp, Yin:2025kls}, completes the \gls{GSN} formalism for gravitational radiation, providing an equivalent analytical prescription for linear perturbations of Kerr \glspl{BH} that has clear numerical advantages over the conventional Teukolsky formalism.

\hfill\begin{acknowledgments}
\emph{Acknowledgments}.---The Center of Gravity is a Center of Excellence funded by the Danish National Research Foundation under Grant No.~DNRF184.
This work was supported by the research Grants No.~VIL37766 and No.~VIL53101 from Villum Fonden, and the DNRF Chair program Grant No.~DNRF162 by the Danish National Research Foundation.
This work has received funding from the European Union's Horizon 2020 research and innovation program under the Marie Sklodowska-Curie Grant Agreement No.~101131233.
The authors would also like to thank Vitor Cardoso for the discussion, Lorena Maga\~na Zertuche for her help with \texttt{qnmfits}, Ariadna Ribes Metidieri for her comments and suggestions, and Manu Srivastava for providing the Mathematica notebook used in his work for crosschecking our radial infall result.
\end{acknowledgments}

\section*{Data Availability}
\label{sec:data_availability}
The data that support the findings of this article are openly available \cite{10.5281/zenodo.18977224}.

\bibliography{BHPT}%

\end{document}


\title{Supplemental Material}
\author{Rico K.~L. Lo}
\affiliation{%
Center of Gravity, Niels Bohr Institute, Blegdamsvej 17, 2100 Copenhagen, Denmark}%
\author{Yucheng Yin}
\affiliation{%
Center of Gravity, Niels Bohr Institute, Blegdamsvej 17, 2100 Copenhagen, Denmark}%
\affiliation{Department of Astronomy, School of Physics, Peking University, 100871 Beijing, China}%
\affiliation{Kavli Institute for Astronomy and Astrophysics at Peking University, 100871 Beijing, China}%


\maketitle

\setcounter{equation}{0}
\setcounter{figure}{0}
\setcounter{table}{0}
\renewcommand{\theequation}{S\arabic{equation}}
\renewcommand{\thefigure}{S\arabic{figure}}
\renewcommand{\thetable}{S\arabic{table}}
\renewcommand{\thesection}{S.\arabic{section}}

\glsaddall
\glsunsetall
\glsreset{RW}

\section{Derivation of the source term for gravitational perturbations in the GSN formalism}\label{sec:derivation}
We begin the derivation by first considering an intermediate variable $\mathcal{X}$, which is related to the \gls{GSN} variable $X$ by \cite{M-Hughes:2000pf, M-Lo:2023fvv}
\begin{equation}
	X(r) = \mathcal{X}(r) \sqrt{\left(r^2 + a^2 \right)\Delta^{s}}.
\end{equation}
Note that we have dropped all the $s\ell m \omega$ subscripts when there is no risk of confusion.
This variable $\mathcal{X}$ satisfies a \gls{RW}-like equation\footnote{This equation is said to be \gls{RW}-like as it reduces to the \gls{RW} equation when $a = 0$.} given by
\begin{equation}
	\Delta^{-s} \left( \Delta^{s+1} \mathcal{X}' \right)' - \Delta F_{1} \mathcal{X}' - U_{1} \mathcal{X} = \mathscr{S},
\end{equation}
where a prime denotes a derivative with respect to $r$, $\mathscr{S}$ is the source term for the $\mathcal{X}$ variable and the expressions for $F_{1}$ and $U_{1}$ can be found in Refs.~\cite{M-Hughes:2000pf, M-Lo:2023fvv}.
We use this equation to write $\mathcal{X}''$ in terms of $\mathcal{X}$ and $\mathcal{X}'$, which is
\begin{equation}
	\mathcal{X}'' = \dfrac{\mathscr{S}}{\Delta} + \dfrac{U_1}{\Delta} \mathcal{X}	+ \left[ F_1 - (s+1) \dfrac{\Delta'}{\Delta} \right] \mathcal{X}'.
\end{equation}

For $s = +2$, we \emph{modify} the inverse transformation from \gls{GSN} solutions $X$ to Teukolsky solutions $R$ the same way for $s = -2$ in Refs.~\cite{M-SASAKI198268, M-10.1143/PTP.67.1788}, as
\begin{equation}
\label{eq:inhomo_X_to_inhomo_R_using_alphabeta}
	R = \dfrac{1}{\eta} \left[ \left( \alpha + \beta' \Delta^{s+1} \right)\mathcal{X} - \beta \Delta^{s+1} \mathcal{X}' \right] + \dfrac{\mathscr{S}}{\eta},
\end{equation}
where again the expressions for $\alpha$, $\beta$ and $\eta$ can be found in Refs.~\cite{M-Hughes:2000pf, M-Lo:2023fvv}.
We then substitute this back into the inhomogeneous radial Teukolsky equation in Eq.~\eqref{M-eq:inhomogeneous_radialteukolsky}, which gives
\begin{multline}
\label{eq:inhomogeneous_radial_teukolsky_like_eqn}
	\Delta^{-s} \left[ \Delta^{s+1} \left( \dfrac{\mathscr{S}}{\eta} \right)' \right]' + \Delta^{-s} \left[ -\beta \Delta^{2s+1} \left(\dfrac{\mathscr{S}}{\eta}\right) \right]' \\ + \left( \alpha - V_{\rm T}\right) \dfrac{\mathscr{S}}{\eta} = -\mathcal{T}.
\end{multline}

The derivation up to this point applies to both $s = \pm 2$. For $s = -2$, refer to Appendix A of Ref.~\cite{M-Yin:2025kls} for a complete proof.
As for $s = +2$, note that we can rewrite Eq.~\eqref{eq:inhomogeneous_radial_teukolsky_like_eqn} into
\begin{equation}
	\mathscr{J} \left[ \mathscr{J} \left( r^2 \Delta \dfrac{\mathscr{S}}{\eta} \right) \right] = - r^2 \mathcal{T},
\end{equation}
where $\mathscr{J} \equiv \partial_r - iK/\Delta$ is a differential operator.\footnote{This $\mathscr{J}$ operator is identical to the $J_{-}$ operator defined in Ref.~\cite{M-Lo:2023fvv}.}
If we introduce an auxiliary variable $\mathcal{W}$ such that
\begin{equation}
	\mathcal{W}(r) = f(r) \exp \left( - \int^{r} i\dfrac{K}{\Delta} d\tilde{r} \right),
\end{equation}
where $f(r)$ is any differentiable function, then $\mathcal{W}'$ can be written as
\begin{equation}
\label{eq:identity_with_Wprime_and_J}
	\mathcal{W}'(r) = \exp \left( - \int^{r} i\dfrac{K}{\Delta} d\tilde{r} \right) \mathscr{J} \left[ f(r) \right].
\end{equation}
Therefore, if we define
\begin{equation}
	\mathcal{W}(r) = r^2 \Delta \dfrac{\mathscr{S}}{\eta} \exp \left( - \int^{r} i\dfrac{K}{\Delta} d\tilde{r} \right),
\end{equation}
then by using the identity in Eq.~\eqref{eq:identity_with_Wprime_and_J} twice, we have
\begin{equation}
	\left(\mathcal{W}' \right)' = \exp \left( -\int^{r} i\dfrac{K}{\Delta} d\tilde{r} \right) \mathscr{J} \left[ \mathscr{J} \left( r^2 \Delta \dfrac{\mathscr{S}}{\eta} \right) \right].
\end{equation}
This means that
\begin{equation}
	\mathcal{W}'' = - r^2 \mathcal{T}  \exp \left( -\int^{r} i\dfrac{K}{\Delta} d\tilde{r} \right),
\end{equation}
which is the \gls{ODE} that one needs to solve to obtain $\mathscr{S}$ from $\mathcal{T}$ for $s = +2$.

Given the source term $\mathscr{S}$ for the variable $\mathcal{X}$, we can convert that into the source term for the \gls{GSN} equation $\mathcal{S}$ simply with
\begin{equation}
	\mathcal{S} = \dfrac{\Delta^{(s+2)/2}}{(r^2 + a^2)^{3/2}} \mathscr{S}.
\end{equation}
Specifically for $s=+2$, we have
\begin{equation}
\label{eq:S_from_calW_for_sp2}
	\mathcal{S} = \dfrac{\eta \Delta \mathcal{W}}{ (r^2 + a^2)^{3/2} r^2}  \exp \left( \int^{r} i\dfrac{K}{\Delta} d\tilde{r} \right).
\end{equation}
This has the same \emph{form} as the source term for $s = -2$, except for having the plus sign inside the exponential function.
In fact,
\begin{equation}
\label{eq:mathcalS_expression}
	{_{\pm2}}\mathcal{S} = \dfrac{\eta \Delta}{ (r^2 + a^2)^{3/2} r^2}  {_{\pm2}}\mathcal{W} \exp \left( \pm \int^{r} i\dfrac{K}{\Delta} d\tilde{r} \right).
\end{equation}
and
\begin{equation}
\label{eq:mathcalW_expression}
    \dfrac{\diff^2 {_{\pm2}}\mathcal{W}}{\diff r^2} = - \frac{r^2}{\Delta^{(2\mp2)/2}} {_{\pm 2}}\mathcal{T}  \exp \left( \mp\int^{r} i\dfrac{K}{\Delta} d\tilde{r} \right).
\end{equation}

\section{Source terms for point particles on arbitrary motions}\label{sec:teukolsky_source_term_for_pp}
We first start with the Teukolsky source term of a point particle for $s = +2$ (or equivalently $\psi_0$), which can be written as \cite{M-Srivastava:2021uku}
\begin{equation}
\label{eq:Teukolsky_sp2_srcterm}
\begin{aligned}
{}_{+2}\mathcal{T}_{\ell m \omega}(r)  & = \mu \int_{\gamma} d\tau\,e^{i\omega t(\tau) - im\varphi(\tau)} \\
& \;\;\;\left\{ \left( A_{ll0} + A_{lm0} + A_{mm0} \right) \delta \left( r - r(\tau) \right) \right. \\
& \;\;\;+ \left[ \left( A_{lm1} + A_{mm1} \right)  \delta \left( r - r(\tau) \right) \right]' \\
& \;\;\;+ \left. \left[ \left( A_{mm2} \right) \delta \left( r - r(\tau) \right) \right]'' \right\},
\end{aligned}
\end{equation}
where $\tau$ is an affine parameter (such as the proper time) parameterizing the trajectory $\gamma$ of the point particle.

Specifically, the $A$ terms above are given by
\begin{subequations}
    \begin{align}
    A_{ll0}&=2\mathscr{A}\bar{\rho}^2\rho\mathscr{L}_1\left[\rho^{-4}\mathscr{L}_2\left(\rho^3S\right)\right]\mathcal{L}^2,\\
        A_{lm0}&=4\mathscr{A}\bar{\rho}^2\left[\mathscr{L}_2S\left(i\frac{K}{\Delta}+\rho+\bar{\rho}\right)\right.\notag\\
        &\left.\qquad-a\sin\theta\frac{K}{\Delta}\left(\rho-\bar{\rho}\right)\right]\mathcal{L}\mathcal{M},\\
        A_{mm0}&=-2\mathscr{A}S\bar{\rho}^2\left[\frac{K^2}{\Delta^2}-2i\rho\frac{K}{\Delta}-i\left(\frac{K}{\Delta}\right)'\right]\mathcal{M}^2,\\
        A_{lm1}&=-4\mathscr{A}\bar{\rho}^2\left[\mathscr{L}_2S+ia\sin\theta(\rho-\bar{\rho})S\right]\mathcal{L}\mathcal{M},\\
        A_{mm1}&=-4\mathscr{A}\bar{\rho}^2S\left(i\frac{K}{\Delta}+\rho\right)\mathcal{M}^2,\\
        A_{mm2}&=2\mathscr{A}\bar{\rho}^2S\mathcal{M}^2,
    \end{align}
\end{subequations}
where $\mathscr{L}_{s} \equiv \partial_\theta + m/\sin\theta - a\omega\sin\theta+s\cot\theta$ is a differential operator on the angular sector, $\rho = - (r - ia \cos \theta)^{-1}$, an overhead bar denotes its complex conjugate, and $\mathscr{A} = -1$ for our choice of normalization conventions (cf Appendix B 1 of Ref.~\cite{M-Yin:2025kls}).\footnote{For example, $\mathscr{A} = -1/\left(2\pi\right)$ for the normalization conventions chosen by Ref.~\cite{M-Srivastava:2021uku}.}
We have also defined some useful combinations of the components of the particle's four velocity $\vec{u}$ where
\begin{subequations}
\begin{align}
        \mathcal{L} & = u^t-a\sin^2\theta u^\varphi-\frac{\Sigma}{\Delta}u^r,\\
        \mathcal{M} & =  -ia\sin\theta u^t+i(r^2+a^2)\sin\theta u^\varphi+\Sigma u^\theta.
\end{align}
\end{subequations}

The corresponding \gls{GSN} source term for $s = +2$, valid for \emph{arbitrary} motions, can be obtained by solving Eqs.~\eqref{eq:mathcalS_expression} and~\eqref{eq:mathcalW_expression} with the ${}_{+2}\mathcal{T}$ given above.

\section{Asymptotic behaviors and boundary conditions for the GSN source term}
Recall that a \gls{GSN} solution ${}_{+2}X$ is related to its corresponding Teukolsky solution ${}_{+2}R$ by [contrast this with Eq.~\eqref{eq:inhomo_X_to_inhomo_R_using_alphabeta}]
\begin{equation}
\label{eq:inhomo_R_to_inhomo_X}
	{}_{+2}R(r) = {}_{+2}\Lambda^{-1} \left( {}_{+2}X \right) + \dfrac{1}{\eta} \dfrac{\left(r^2 + a^2\right)^{3/2}}{\Delta^{2 }} {}_{+2}\mathcal{S},
\end{equation}
and that we impose a purely ingoing boundary condition for the inhomogeneous solution ${}_{+2}X^{\rm inhomo}$ at the horizon that
${}_{+2}X^{\rm inhomo}(\rs \to -\infty) = {}_{+2}X^{\hor} e^{-i\kappa \rs}$.
Note that the inverse operator ${}_{+2}\Lambda^{-1}$ transforms a function that goes like $e^{-i\kappa \rs}$ into
\begin{equation}
    {}_{+2}\Lambda^{-1} \left( e^{-i\kappa \rs} \right) \propto \dfrac{1}{\Delta^{2}} e^{-i\kappa \rs},
\end{equation}
which has the correct asymptotic behavior needed for the corresponding inhomogeneous Teukolsky solution that ${}_{+2}R^{\rm inhomo}(r \to \rp) = {}_{+2}R^{\hor} \Delta^{-2}  e^{-i\kappa \rs}$ [cf Eq.~\eqref{M-eq:R_inhomo}] \cite{M-Lo:2023fvv}.

Therefore, if we want the asymptotic amplitudes ${}_{+2}R^{\hor}$ and ${}_{+2}X^{\hor}$ to be off only by a conversion factor ${\left(B^{\rm trans}_{\rm T}/B^{\rm trans}_{\rm SN}\right)}$, then we need to require the contribution near the horizon from the second term in Eq.~\eqref{eq:inhomo_R_to_inhomo_X} to be subdominant compared to the first term, such that it vanishes at the horizon.
Concretely, it means that the \gls{GSN} source term ${}_{+2}\mathcal{S}$ (specifically, its modulus) needs to satisfy
\begin{equation}
    \left| {}_{+2}\mathcal{S}(r \to \rp) \right| \sim \Delta^{n},\;n \geq 1.
\end{equation}
Using Eq.~\eqref{eq:mathcalS_expression}, this condition implies that the auxiliary variable $\mathcal{W}$ should satisfy
\begin{equation}
\label{eq:reg_bc_for_mathcalW}
    \left| {}_{+2}\mathcal{W}(r \to \rp) \right| \sim \Delta^{n},\;n \geq 0,
\end{equation}
meaning that ${}_{+2}\mathcal{W}$ should be regular at the \gls{BH} horizon, though it can be oscillatory.

In order to understand the asymptotic behavior of ${}_{+2}\mathcal{W}$ near the \gls{BH} horizon, we first inspect the Teukolsky source term ${}_{+2}\mathcal{T}$ in Eq.~\eqref{eq:Teukolsky_sp2_srcterm}.
We see that the Teukolsky source term near the \gls{BH} horizon diverges at most $\sim 1/\Delta^2$.
Thus, we expand the source term ${}_{+2}\mathcal{T}$ near $r = \rp$ as
\begin{equation}
\label{eq:teukolsky_src_term_near_hor}
    {}_{+2}\mathcal{T}(r) \approx T_{-2} \left( r - \rp\right)^{-2} + T_{-1} \left( r - \rp\right)^{-1} + T_{0} + \cdots,
\end{equation}
where $T_{-2,-1,0,\cdots}$ are expansion coefficients independent of $r$.
Near the \gls{BH} horizon, we have
\begin{equation}
    \exp \left( -\int^{r} i\dfrac{K}{\Delta} d\tilde{r} \right) = e^{-i\omega r} \sum_{j=0}^{\infty} a_j \left( r - \rp \right)^{j + iq},
\end{equation}
where $q = \left(a m - 2 \rp \omega \right)/\left(\rp - \rminus \right)$ and $a_{0,1,\cdots}$ are again some expansion coefficients independent of $r$.
Putting all these into the \gls{ODE} in Eq.~\eqref{eq:mathcalW_expression}, we have
\begin{equation}
\label{eq:Wpp_near_horizon}
    {{_{+2}}\mathcal{W}}'' \approx - a_0 T_{-2} r^2 \left( r - \rp \right)^{-2+iq} e^{-i\omega r} + \cdots.
\end{equation}

To show that the regular boundary condition in Eq.~\eqref{eq:reg_bc_for_mathcalW} is satisfied, we consider the most singular piece in Eq.~\eqref{eq:Wpp_near_horizon}, which is exactly the term that is shown explicitly.
Notice that it can be integrated analytically using incomplete gamma functions.
We can then evaluate the particular solution ${}_{+2}\mathcal{W}^{\rm part}$ at the \gls{BH} horizon as
\begin{multline}
    {}_{+2}\mathcal{W}^{\rm part}(r = \rp) = -\dfrac{a_{0}T_{-2} \left( -i \omega \right)^{-iq} e^{i\omega \rp}}{\omega^2} \\
    \left[ 2i \rp \omega \Gamma\left(1+iq\right) - \Gamma\left(2+iq\right) + \rp^2 \omega^2 \Gamma\left(iq\right) \right],
\end{multline}
and its derivative ${{}_{+2}\mathcal{W}^{\rm part}}'$ at the horizon as
\begin{multline}
    {{}_{+2}\mathcal{W}^{\rm part}}'(r = \rp) = -a_{0}T_{-2} \left( -i \omega \right)^{-1-iq} e^{i\omega \rp} \\
    \left[ 2i \rp \omega \Gamma\left(iq\right) - \Gamma\left(1+iq\right) + \rp^2 \omega^2 \Gamma\left(-1+iq\right) \right],
\end{multline}
where $\Gamma(z)$ is the gamma function, and they are indeed finite for $\omega \neq 0$.

Combining the particular solution with the complementary solution ${}_{+2}\mathcal{W}^{\rm comp}(r) = w_0 + w_1(r - \rp)$, the general solution near the \gls{BH} horizon is given by
\begin{equation}
    {}_{+2}\mathcal{W}(r \to \rp) = w_0 + w_1(r - \rp) + \mathcal{W}^{\rm part}(r \to \rp).
\end{equation}
It is now easy to show that the convolution integral in Eq.~\eqref{M-eq:full_inhomo_X_using_green_func} for ${}_{+2}X^{\hor}$ in the \gls{GSN} formalism is absolutely convergent.
This is because, in the worst-case scenario, the integrand 
\begin{equation}
\label{eq:convolution_integrand_gsn}
    \left|X^{\rm up}(\rs) \mathcal{S}(\rs)/\eta(\rs)\right|_{\rs \to -\infty} \sim \Delta \to 0,
\end{equation}
since both $|X^{\rm up}(\rs \to -\infty)|$ and $|\eta(\rs \to -\infty)|$ scale like $\bigO(1)$.
Near spatial infinity, the complementary solution $w_{1}\left(r - \rp \right)$ dominates $\mathcal{W}$, and thus $|\mathcal{S}(r \to \infty)| \to 0$.
The integrand goes to zero, i.e.,
\begin{equation}
    \left|X^{\rm up}(\rs) \mathcal{S}(\rs)/\eta(\rs)\right|_{\rs \to \infty} \to 0,
\end{equation}
since again both $|X^{\rm up}(\rs \to \infty)|$ and $|\eta(\rs \to \infty)|$ scale like $\bigO(1)$.
Meanwhile, the convolution integral in the Teukolsky formalism [cf Eq.~\eqref{M-eq:R_inhomo}] is divergent, since the integral
\begin{equation}
\label{eq:convolution_integrand_teuk_div}
    \left|R^{\rm up}(r) \Delta^{2} \mathcal{T}(r)\right|_{r \to \rp} \sim \Delta^{-2} \to \infty.
\end{equation}

\section{Analytical GSN source term for ultrarelativistic radial infalls}
\label{sec:ultrarelativistic_radial_infall_src_term}
Here we show that it is possible to have an analytical \gls{GSN} source term, for instance, if a point particle is falling radially into a nonrotating \gls{BH}, i.e., $a = 0$, with an ultrarelativistic initial speed/Lorentz boost factor.

For radial motions ($\theta = 0$), we only need to consider modes with $m = 0$ without loss of generality.
For ultrarelativistic radial infalls starting from infinity, $\diff t/\diff r$ is approximately
\begin{equation}
    \dfrac{\diff t}{\diff r} = \dfrac{\diff t/\diff \tau}{\diff r/\diff \tau} \approx -\dfrac{r}{r-2}.
\end{equation}
Therefore, we have $t(r) \approx -\rs(r)$. The Teukolsky source term, using Eq.~\eqref{eq:Teukolsky_sp2_srcterm}, is just
\begin{equation}\label{Eq.Teukolsky_Source_RadialInfall}
    {}_{+2}\mathcal{T}(r) = \mu \dfrac{A_{ll0}(r)}{u^r(r)} e^{-i\omega\rs} = -32  \left.\dfrac{\partial^2 S}{\partial \theta^2}\right|_{\theta = 0} \mathcal{E}\mu \dfrac{e^{-i\omega\rs}}{(r-2)^2},
\end{equation}
where we used the fact that
\begin{equation}
    \mathscr{L}_1\left[\rho^{-4}\mathscr{L}_2\left(\rho^3S\right)\right] = -4 \left. \dfrac{\partial^2 {}_{2}S_{\ell 0 \omega}(\theta, 0)}{\partial {\theta}^2}  \right|_{\theta = 0} r.
\end{equation}
From here, we see that the Teukolsky source term goes like $1/(r-\rp)^2$ as $r \to \rp = 2$.

The \gls{ODE} for $\mathcal{W}$ in Eq.~\eqref{eq:mathcalW_expression} can thus be written as
\begin{equation}
\label{eq:ODE_for_mathcalW_ultrarelativistic}
    {\mathcal{W}}'' = 32 \left.\dfrac{\partial^2 S}{\partial {\theta}^2}\right|_{\theta = 0} \mathcal{E}\mu \dfrac{r^2}{(r-2)^2} e^{-2i\omega\rs}. 
\end{equation}
Recall that the tortoise coordinate $\rs$ (for $r > 2$) when $a = 0$ is just
\begin{equation}
    \rs = r + 2\ln \left( \dfrac{r - 2}{2} \right).
\end{equation}
This means that $e^{-2i\omega \rs}$ can also be written as
\begin{equation}
e^{-2i\omega \rs} = e^{-2i\omega r} \left(\dfrac{r - 2}{2}\right)^{-4i\omega}.
\end{equation}
Equation~\eqref{eq:ODE_for_mathcalW_ultrarelativistic} is then
\begin{equation}
{\mathcal{W}}'' = C r^2 (r - 2)^{-2-4i\omega} e^{-2i\omega r},
\end{equation}
where $C = 2^{5 + 4i\omega} \left. \dfrac{\partial^2 {}_{2}S_{\ell 0 \omega}(\theta, 0)}{\partial {\theta}^2}  \right|_{\theta = 0} \mathcal{E}\mu$.
This \gls{ODE} can be solved using incomplete gamma functions.
Specifically, the solution $\mathcal{W}$ is given by
\begin{multline}
\label{eq:calWll_radialinfall}
    \mathcal{W}(r) = -\dfrac{C e^{\left( 2i\omega - 1 \right)\ln 4 -4i\omega\left[1 - \ln(i\omega)\right]}}{\omega^2 \left( 1 + 4i\omega \right)}  \\
    \Bigl\{ \left( 1 + 8i\omega \right) e^{-4i\omega[\ln (4i\omega)-1]} e^{-2i\omega \rs} \\ + \left( 4i\omega - 8r\omega^2 \right)  \gamma\left(-4i\omega, 2i\omega(r-2)\right) \Bigl \},
\end{multline}
where $\gamma(a,z) \equiv \int_{0}^{z} t^{a-1} e^{-t} dt$ is the lower incomplete gamma function.

\begin{figure}[!htb]
\includegraphics[width=\columnwidth]{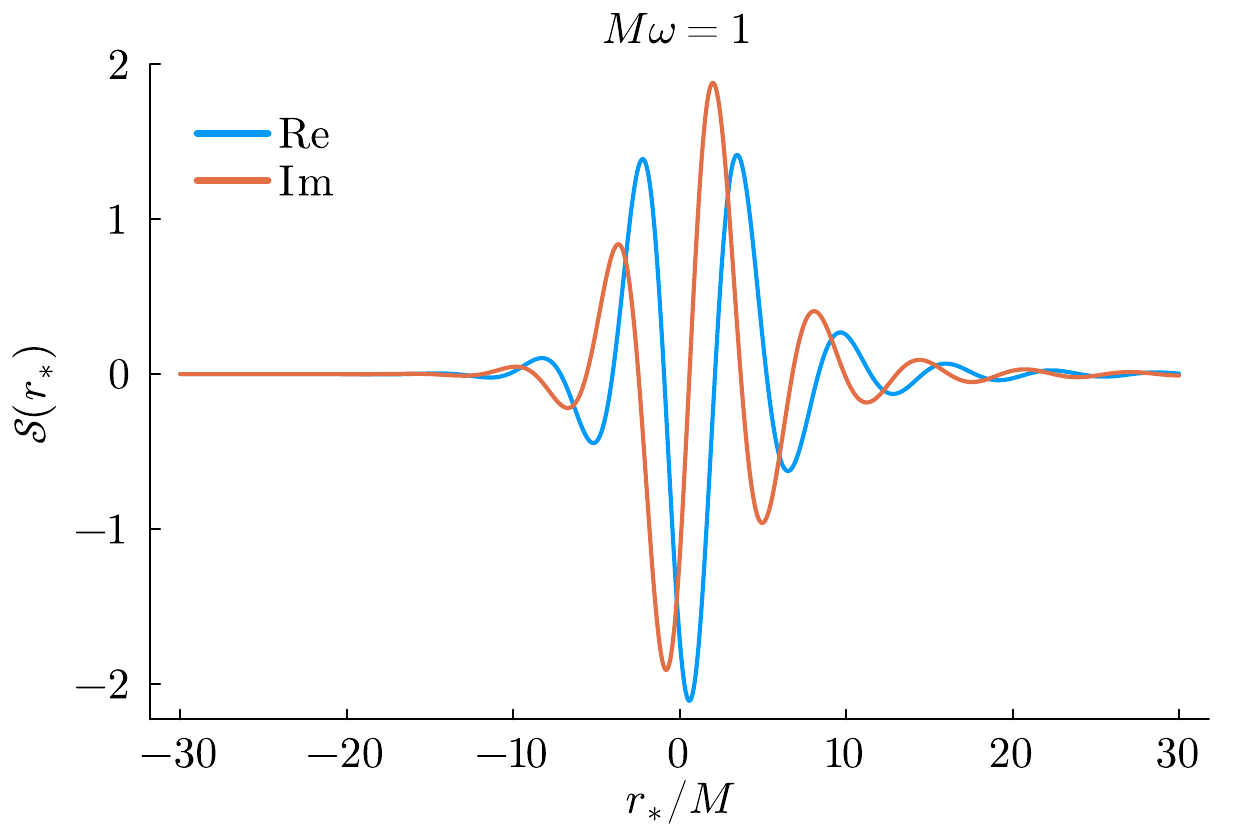}
\caption{\label{fig:radial-infall-srcterm-sp2}The \gls{GSN} source term $\mathcal{S}$ for $s = +2$ with an ultrarelativistic particle falling into a nonrotating \gls{BH} radially. The source term is nonzero at around $\rs \sim 0M$ and decays to zero quickly.}
\end{figure}

The \gls{GSN} source term $\mathcal{S}$ can be obtained by substituting Eq.~\eqref{eq:calWll_radialinfall} into Eq.~\eqref{eq:S_from_calW_for_sp2}, which is also shown in Fig.~\ref{fig:radial-infall-srcterm-sp2}. The source term $\mathcal{S}(\rs)$ here is indeed short-ranged, just like its $s=-2$ counterpart \cite{M-Cardoso:2002jr}, that it is nonzero at around $\rs \sim 0M$ and then decays to zero quickly.

\begin{figure}[htbp!]
\begin{centering}
\subfloat[]{\includegraphics[width=\columnwidth]{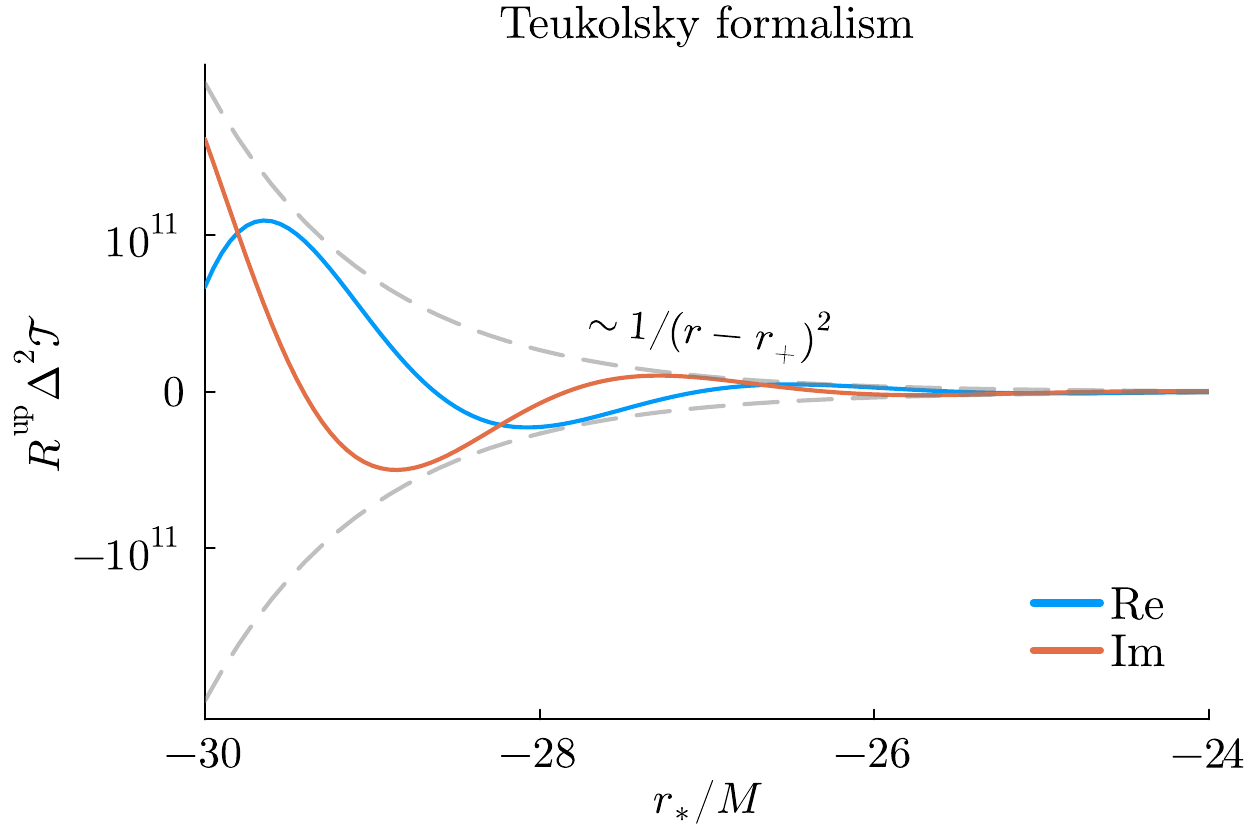}}\\
\subfloat[]{\includegraphics[width=\columnwidth]{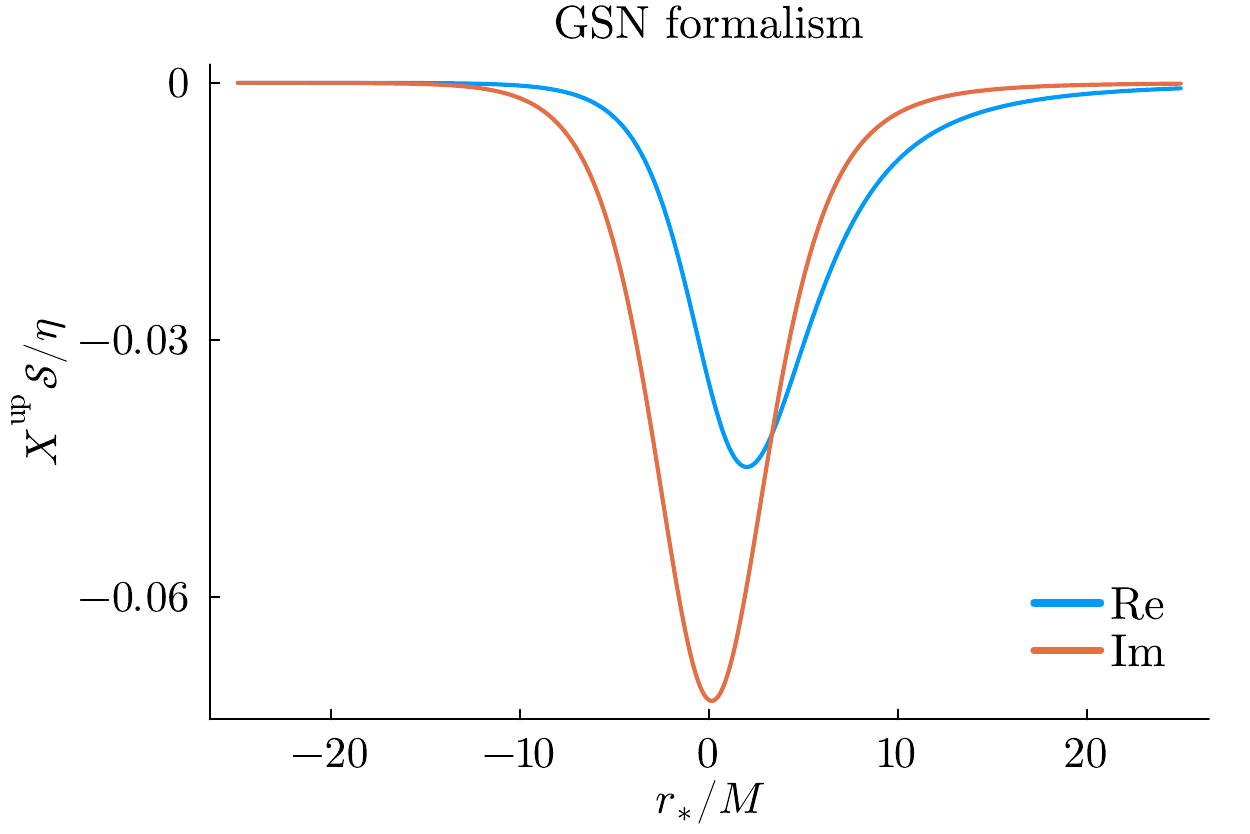}}
\end{centering}
\caption{\label{fig:convolution_integrand}The integrand in the convolution integral when $M\omega = 1$ under the Teukolsky formalism (upper) and the \gls{GSN} formalism (lower) for the radial infall of an ultrarelativistic particle into a nonspinning \gls{BH}, respectively. We see that the convolution integrand under the Teukolsky formalism diverges like $1/(r - \rp)^2$ indicated by the dash lines, while that under the \gls{GSN} formalism is finite.}
\end{figure}

Figure~\ref{fig:convolution_integrand} shows the asymptotic behaviors (near $r \to \rp$ or $\rs \to -\infty$) of the integrands in the convolution integrals under the Teukolsky (upper panel) and the \gls{GSN} (lower panel) formalism for the ultrarelativistic radial infall, respectively.
We see that the integrand under the Teukolsky formalism, $R^{\rm up}\Delta^2 \mathcal{T}$, diverges like $1/\left( r - \rp\right)^2$ [cf Eq.~\eqref{eq:convolution_integrand_teuk_div}].
On the other hand, the integrand under the \gls{GSN} formalism, $X^{\rm up}\mathcal{S}/\eta$, is finite [cf Eq.~\eqref{eq:convolution_integrand_gsn}] and localized near $\rs \sim 0M$.
This means that when evaluating the convolution integral numerically, one can choose smaller values for the numerical lower and upper limits.

\section{Regularization of the Teukolsky source term for ultrarelativistic radial infalls}
The source terms and thus the Green's function integrals in the Teukolsky formalism require some regularization when the source terms do not vanish at the \gls{BH} horizon.
Here, we follow the regularization procedures in Ref.~\cite{M-Srivastava:2021uku} in order to benchmark and compare the solutions from the two formalisms.

Recall that an inhomogeneous solution to the radial Teukolsky equation for $s = +2$ can be constructed using the Green's function method as
\begin{multline}
    {}_{+2}R^{\mathrm{inhomo}}(r \to \rp) = \\ \frac{{}_{+2}R^{\mathrm{in}}_{\ell m \omega} (r)}{W_{R}}\int_{\rp}^\infty {}_{+2}R_{\ell m\omega}^{\rm up} (r)\Delta^2{}_{+2}\mathcal{T}_{\ell m\omega}(r)\diff r,
\end{multline}
where ${}_{+2}\mathcal{T}_{\ell m\omega}(r)$ is the source term given in Eq.~\eqref{Eq.Teukolsky_Source_RadialInfall}.
The integrand is divergent at the horizon as shown in Fig.~\ref{fig:convolution_integrand}. Following Ref.~\cite{M-Srivastava:2021uku}, one possible solution to regularize the integral is to construct an ansatz
\begin{equation}
    R_{\mathrm{A}}(r)=\left(\frac{A_0}{r\Delta^2}+\frac{A_1}{r^3\Delta}\right)e^{i\omega t(r)-im\varphi(r)},
\end{equation}
where, in this case,
\begin{subequations}
    \begin{align}
        A_0&=-\frac{8C(4+\lambda-28i\omega)}{2^{4i\omega}B},\\
        A_1&=-\frac{8C(12+3\lambda-32i\omega)}{2^{4i\omega}B},\\
        B&=24+10\lambda+\lambda^2-132i\omega-32i\lambda\omega-320\omega^2,
    \end{align}
\end{subequations}
and $\lambda = \ell \left(\ell + 1 \right) - 6$ is a separation constant from the angular Teukolsky equation (see, for example, Appendix A of Ref.~\cite{M-Lo:2023fvv}).
Note that these expressions are simple only because the Teukolsky source term is exactly $1/(r-\rp)^2$ multiplied by $e^{i\omega t(r)}$ and some constants [cf Eq.~\eqref{Eq.Teukolsky_Source_RadialInfall}].

The regularized source term ${}_{+2}\tilde{\mathcal{T}}$ is then constructed as
\begin{equation}
\begin{aligned}
    {}_{+2}\tilde{\mathcal{T}}(r) & = {}_{+2}\mathcal{T}(r)-\frac{{}_{+2}\mathcal{D}\left[R_{\mathrm{A}}(r)\right]}{\Delta^2}\\
    & =\frac{Ce^{-i\omega\rs}}{2^{4i\omega}r^5B} \bigl\{ -320r^2\omega^2(r+4)-2048i\omega \\
    & \;\;\;\; -4i\omega r[r(4+\lambda)(33+8\lambda)+4(35+8\lambda)]\\
    & \;\;\;\; +(4+\lambda)[192+r(r^2+4r+12)(6+\lambda)] \bigl\},
\end{aligned}
\end{equation}
where ${}_{+2}\mathcal{D}$ is the differential operator representing the radial Teukolsky equation. We can clearly see that the new source term ${}_{+2}\tilde{\mathcal{T}}(r\to\rp)$ is now $\sim\bigO(1)$, making the Green's function integral convergent.
The now-regularized Teukolsky solution can be expressed using a (convergent) convolution integral as
\begin{multline}
    {}_{+2}R_{\ell 0\omega}(r\to\rp)=\left(\frac{A_0}{r\Delta^2}+\frac{A_1}{r^3\Delta}\right)e^{-i\omega \rs}\\
    +\frac{B^{\mathrm{trans}}_{\mathrm{T}} e^{-i\omega\rs}}{W_{R}\Delta^2}\int_{\rp}^\infty {}_{+2}R_{\ell 0\omega}^{\rm up} (r)\Delta^2{}_{+2}\tilde{\mathcal{T}}_{\ell 0\omega}(r)\diff r,
\end{multline}
where we have used the fact that ${}_{+2}R^{\mathrm{in}}_{\ell 0\omega}(r \to \rp) = B^{\mathrm{trans}}_{\mathrm{T}} e^{-i\omega \rs}/\Delta^2$.

\section{Evaluating GSN asymptotic amplitudes using integration-by-parts}
Following the same procedure in our earlier work \cite{M-Yin:2025kls}, we can efficiently evaluate the integral
\begin{equation}
    I \equiv \int_{\rp}^{\infty} {}_{+2}Y^{\rm up}_{\ell m \omega}(r) \dfrac{\diff^2 {}_{+2}\mathcal{W}_{\ell m \omega}}{\diff r^2} dr,
\end{equation}
coming from the integration-by-parts approach [cf Eq.~\eqref{M-eq:XH_lmnk}] by evaluating the Dirac delta functions in the Teukolsky source term [cf Eq.~\eqref{eq:Teukolsky_sp2_srcterm}].
Specifically, the integral $I$ can be written as
\begin{equation}
    \begin{aligned}
    I & = - \mu \int_{\gamma}\left(W_{ll}\mathcal{L}^2+W_{lm}\mathcal{L}\mathcal{M}+W_{mm}\mathcal{M}^2\right)\\
        &\qquad\quad\times e^{i\omega t(\tau)-im\varphi(\tau)} \diff \tau,
    \end{aligned}
\end{equation}
where $W_{ll}$, $W_{lm}$ and $W_{mm}$ are given by
\begin{subequations}
    \begin{align}
        W_{ll} & = 2\mathscr{A}\rho\bar{\rho}^2\mathscr{L}_1\left[\rho^{-4}\mathscr{L}_2\left(\rho^3 S\right)\right]r^2Y\mathrm{phase},\\
        W_{lm} & = 4\mathscr{A}r\bar{\rho}^2\left\{\left(\mathscr{L}_2S\right)\left(\rho+\bar{\rho}\right)rY\right.\\
        &\left.+\left[\mathscr{L}_2 S+i a\sin\theta\left(\rho-\bar{\rho}\right)S\right]\left(2Y+rY'\right)\right\}\mathrm{phase},\notag\\
        W_{mm} & = 4\mathscr{A}S\bar{\rho}^2\biggl[\frac{X}{2\sqrt{r^2+a^2}}+\left(Y+2rY'\right)\mathrm{phase}\\
        &+\rho r\left(2Y+rY'\right)\mathrm{phase}\biggr],\notag\\
        \text{phase} & = \exp\left(-i \int^r\frac{K}{\Delta}\diff\tilde{r}\right) \\
        &=\exp\left(-i \omega \rs+\frac{i am}{2\sqrt{1-a^2}}\ln\frac{r-\rp}{r-\rminus}\right)\notag.
    \end{align}
\end{subequations}

\section{Fluxes towards black hole horizons}
The averaged fluxes towards \gls{BH} horizons are related to the asymptotic \gls{GSN} amplitude ${}_{+2}X^{\hor}_{\ell m n k}$ defined in Eqs.~\eqref{M-eq:full_inhomo_X_using_green_func} and \eqref{M-eq:XH_lmnk}. Using Ref.~\cite{M-Chandrasekhar:1985kt} and the conversion factor $B^{\rm trans}_{\rm T}/B^{\rm trans}_{\rm SN}$ from Ref.~\cite{M-Lo:2023fvv}, we have
\begin{subequations}\label{Eq.EnergyFlux}
    \begin{align}
        \left\langle\dot{\mathcal{E}}\right\rangle^{\hor}_{\ell mnk} & =\frac{\omega\left|{\left(B^{\rm trans}_{\rm T}/B^{\rm trans}_{\rm SN}\right)} {}_{+2}X^{\hor}_{\ell m n k}\right|^2}{64\pi(2\rp)^3\kappa(\kappa^2+4\epsilon_0^2)},\\
        \left\langle\dot{\mathcal{L}}_z\right\rangle^{\hor}_{\ell mnk} & =\frac{m\left|{\left(B^{\rm trans}_{\rm T}/B^{\rm trans}_{\rm SN}\right)} {}_{+2}X^{\hor}_{\ell m n k}\right|^2}{64\pi(2\rp)^3\kappa(\kappa^2+4\epsilon_0^2)},\\
        \left\langle\dot{\mathcal{Q}}\right\rangle^{\hor}_{\ell mnk} & =\frac{\left(\mathcal{L}_{mnk}+k\Upsilon_\theta\right)\left|{\left(B^{\rm trans}_{\rm T}/B^{\rm trans}_{\rm SN}\right)} {}_{+2}X^{\hor}_{\ell m n k}\right|^2}{32\pi(2\rp)^3\kappa(\kappa^2+4\epsilon_0^2)},
    \end{align}
\end{subequations}
where $\epsilon_0=\sqrt{1-a^2}/(4\rp)$ and $\Upsilon_\theta$ is the Mino frequency of the polar motion (cf Sec.~II C 1 of Ref.~\cite{M-Yin:2025kls}).
The $\mathcal{L}_{mnk}$ term is given by
\begin{subequations}
    \begin{align}
        \mathcal{L}_{mnk} & = m\langle\cot^2\theta\rangle\mathcal{L}_z-a^2\omega_{mnk}\langle\cos^2\theta\rangle\mathcal{E},\\
        \langle\cot^2\theta\rangle & =\frac{1}{\pi}\int_0^{\pi}\left[\cot\theta(\phi_\theta)\right]^2\diff\phi_\theta,\\
        \langle\cos^2\theta\rangle & =\frac{1}{\pi}\int_0^{\pi}\left[\cos\theta(\phi_\theta)\right]^2\diff\phi_\theta.
    \end{align}
\end{subequations}

\section{Quasinormal mode fits of the shear waveform from an ultrarelativistic radial infall}
\begin{figure*}[htbp!]
\begin{centering}
\subfloat[the fundamental mode $\mathcal{A}_{200}$]{\includegraphics[width=\columnwidth]{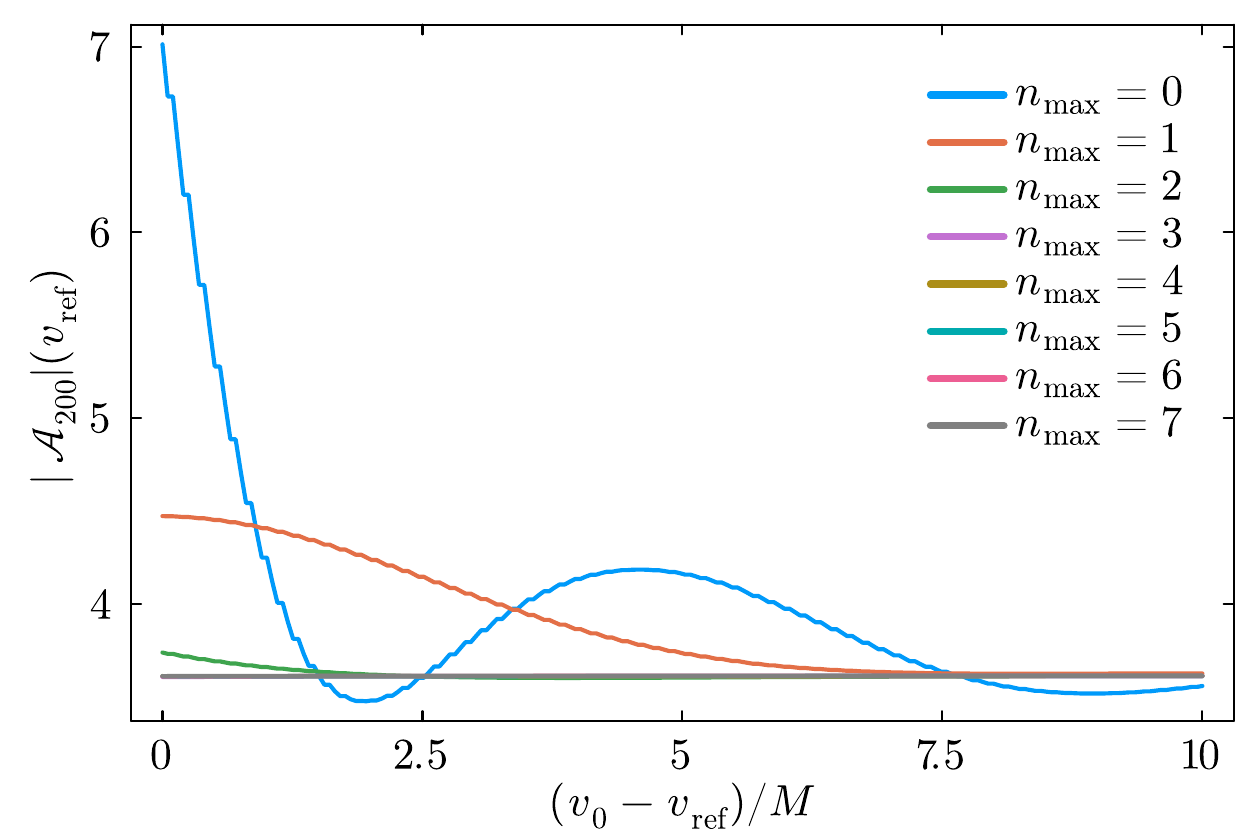}}
\subfloat[the first overtone $\mathcal{A}_{201}$]{\includegraphics[width=\columnwidth]{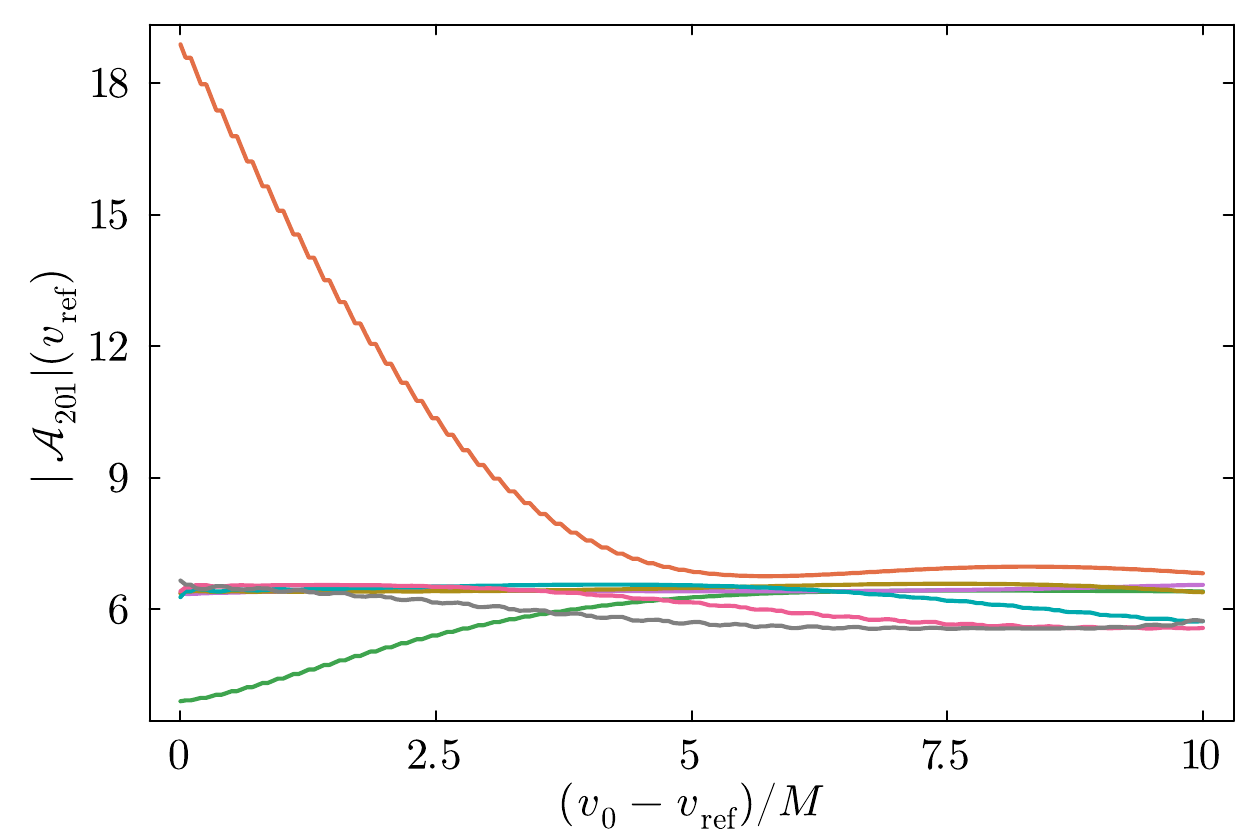}}
\\
\subfloat[the second overtone $\mathcal{A}_{202}$]{\includegraphics[width=\columnwidth]{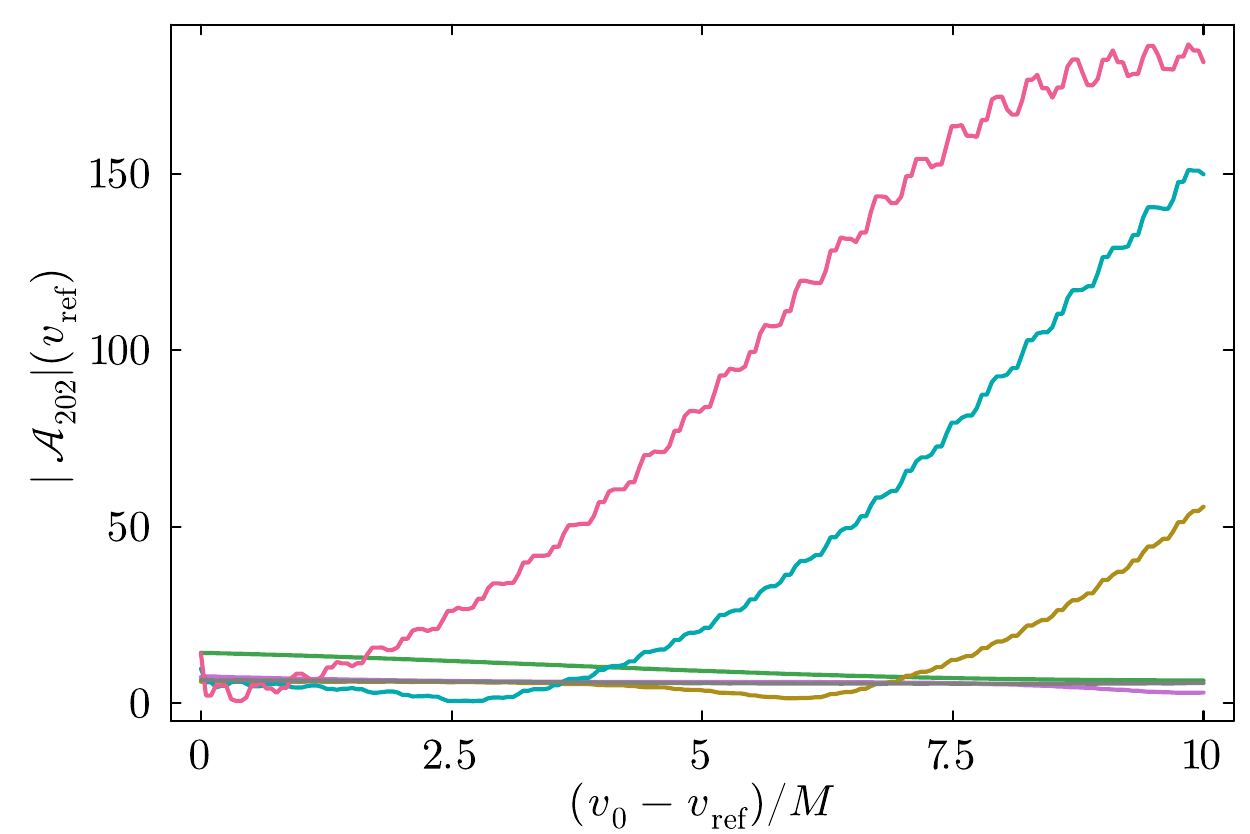
}}
\subfloat[the third overtone $\mathcal{A}_{203}$]{\includegraphics[width=\columnwidth]{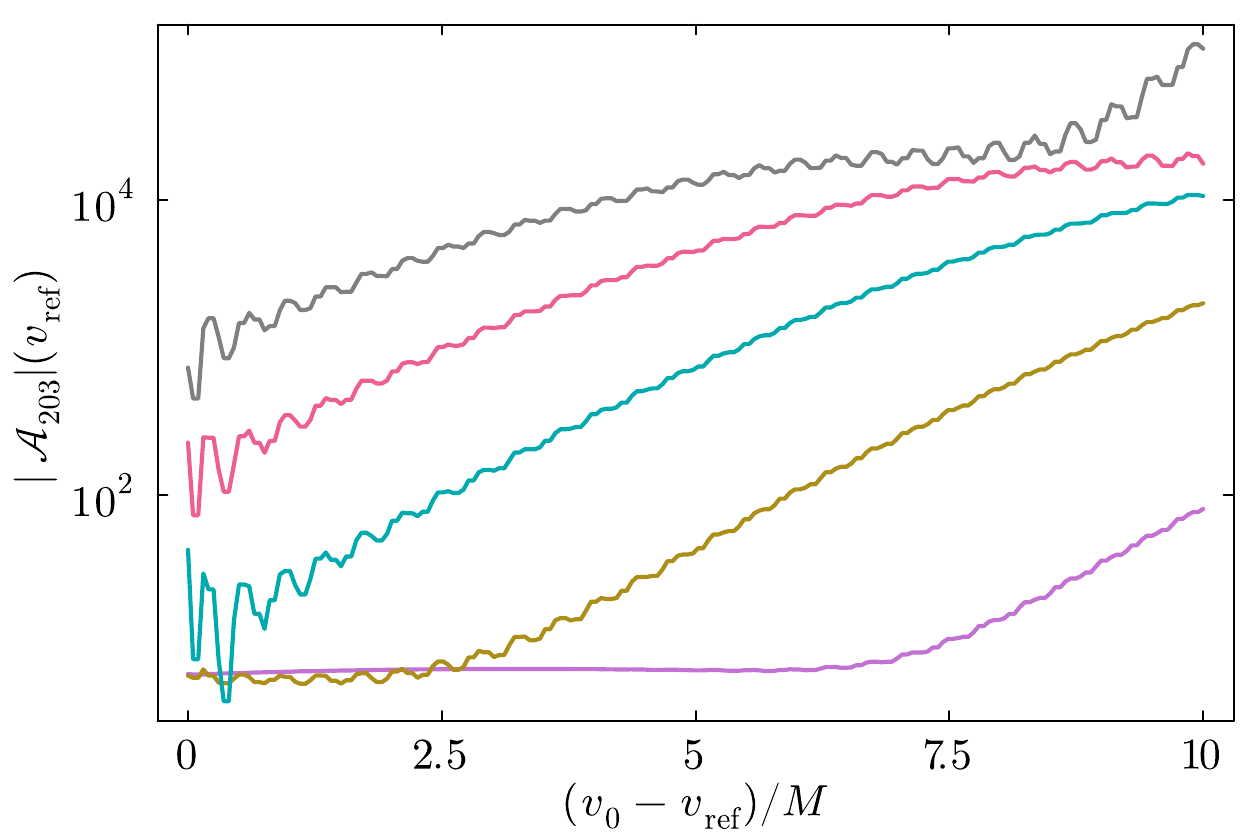
}}\\
\end{centering}
\caption{\label{fig:qnm-fit-fundamental}The fitted \gls{QNM} amplitudes as a function of the starting time $v_0$ of the fits. The reference time $v_{\rm ref}$ is fixed to $0M$. We see that the fundamental mode alone (i.e., $n_{\rm max} = 0$) is insufficient for a stable extraction across different starting times, and that we cannot robustly confirm the presence of \glspl{QNM} beyond the third overtone from our perturbation waveform.}
\end{figure*}

To confirm the presence of \glspl{QNM} in the shear waveform shown in Fig.~1, we perform multiple fits that include different number of overtones (from $n_{\rm max} = 0$ to $n_{\rm max} = 7$), starting from $v_{0} = 0M$ to $v_{0} = 10M$.
Some of the extracted amplitudes are shown in Fig.~\ref{fig:qnm-fit-fundamental}, for the fundamental mode $\mathcal{A}_{200}$ (upper left panel), the first overtone $\mathcal{A}_{201}$ (upper right panel), the second overtone $\mathcal{A}_{202}$ (lower left panel), and the third overtone $\mathcal{A}_{203}$ (lower right panel), respectively.

We see that $\mathcal{A}_{20n}$ for $n \geq 4$ cannot be reliably extracted from our perturbation waveform, in a manner that is stable and robust against different choices of starting times. Furthermore, the fundamental mode alone (i.e., $n_{\rm max} = 0$) is insufficient to fit the waveform, and that overtones are needed for a good fit. In fact, the amplitudes for the first three overtones $\mathcal{A}_{201,202,203}(v_{\rm ref})$ are almost twice as large as that for the fundamental mode $\mathcal{A}_{200}(v_{\rm ref})$, evaluated at the reference time $v_{\rm ref} = 0M$.

\bibliography{BHPT}%